\documentclass[aps,prb,twocolumn,groupedaddress,showpacs,floatfix]{revtex4}
\usepackage{graphicx}
\usepackage{dcolumn}
\usepackage{bm}
\usepackage{stmaryrd}
\usepackage{latexsym}
\usepackage{amssymb}
\usepackage{amsfonts}
\usepackage{amsmath}

\begin{document}

\title{Bethe-Ansatz density-functional theory of ultracold repulsive fermions in one-dimensional optical lattices}
\author{Gao Xianlong}
\author{Marco Polini}
\email{m.polini@sns.it}
\author{M. P. Tosi}
\affiliation{NEST-CNR-INFM and Scuola Normale Superiore, I-56126 Pisa, Italy}
\author{Vivaldo L. Campo, Jr.}
\affiliation{Centro Internacional de F\'{\i}sica da Mat\'eria Condensada,
Universidade de Bras\'{\i}lia,
Caixa Postal 04513, 70919-970 Bras\'{\i}lia, Brazil}
\author{Klaus Capelle}
\affiliation{Departamento de F\'{\i}sica e Inform\'atica,
Instituto de F\'{\i}sica de S\~ao Carlos,
Universidade de S\~ao Paulo,
Caixa Postal 369, 13560-970 S\~ao Carlos, S\~ao Paulo, Brazil}
\author{Marcos Rigol}
\affiliation{Physics Department, University of California, Davis, CA 95616, USA}
\affiliation{Institut f\"ur Theoretische Physik III, Universit\"at Stuttgart, 70550 
Stuttgart, Germany}

\date{\today}
\begin{abstract}
We present an extensive numerical study of ground-state properties of 
confined repulsively interacting fermions on one-dimensional optical lattices. 
Detailed predictions for the atom-density profiles are obtained from 
parallel Kohn-Sham density-functional calculations and quantum Monte Carlo simulations. 
The density-functional calculations employ a Bethe-Ansatz-based local-density 
approximation for the correlation energy, which accounts for Luttinger-liquid 
and Mott-insulator physics. Semi-analytical and fully numerical formulations 
of this approximation are compared with each other and with a cruder 
Thomas-Fermi-like local-density approximation for the total energy. Precise
quantum Monte Carlo simulations are used to assess the reliability of the various 
local-density approximations, and in conjunction with these allow to obtain
a detailed microscopic picture of the consequences of the interplay between
particle-particle interactions and confinement in one-dimensional systems of strongly 
correlated fermions.
\end{abstract}
\maketitle

\section{Introduction}

Strongly correlated one-dimensional ($1D$) quantum liquids and gases are nowadays available in a large number of different laboratory systems ranging from single-wall carbon nanotubes~\cite{Saito_book} to semiconductor nanowires~\cite{Tans_1998}, conducting molecules~\cite{Nitzan_2003}, and trapped atomic gases~\cite{cold_atoms_low_D,paredes_2004,kinoshita_2004,moritz_2005}. Chiral Luttinger liquids at fractional quantum Hall edges~\cite{xLL} also provide a beautiful example of $1D$ conducting quantum liquids and have been the subject of intense experimental~\cite{xLL_experiments} and theoretical efforts~\cite{giovanni_allan}.

There are two fundamental key features that are common to all these $1D$ systems. (i) Independently of statistics, 
their effective low-energy description is based on a harmonic theory of long-wavelength fluctuations~\cite{haldane_harmonic} due to the interplay between topology and interactions. (ii) In the most interesting and exciting experimental situations the translational invariance of the liquid can be broken due to the presence of inhomogeneous external fields of different types, such as magnetic traps in the case of ultracold atomic gases~\cite{bec} or Hall bar constrictions in the case of fractional quantum Hall edges~\cite{xLL_experiments}. These strong perturbations induce the appearance of new length scales causing novel physical behaviors relative to the corresponding unperturbed, translationally invariant model system.

A powerful theoretical tool to study the interplay between interactions and inhomogeneous external fields of arbitrary shape is density-functional theory (DFT)~\cite{wk,d&g,joulbert_1998,Giuliani_and_Vignale}, which is based on the Hohenberg-Kohn theorem~\cite{hk} and on the Kohn-Sham mapping to an auxiliary noninteracting system~\cite{ks}. Many-body effects enter DFT {\it via} the exchange-correlation (xc) functional, which is often treated by the local-density approximation (LDA)~\cite{wk,d&g,joulbert_1998,Giuliani_and_Vignale,hk,ks}. The essence of LDA is to locally approximate the xc energy of the inhomogeneous system under study with that of an interacting homogeneous reference fluid, whose correlations are transferred by the LDA to the inhomogeneous system. For example, for inhomogeneous $2D$ and $3D$ electronic systems the underlying reference fluid is normally the homogeneous electron liquid (EL)~\cite{Pines_and_Nozieres,Giuliani_and_Vignale}, whose xc energy is known to a high degree of numerical precision thanks to the quantum Monte Carlo (QMC) technique~\cite{qmc}. 
However, these ELs are believed to be normal Fermi liquids~\cite{Pines_and_Nozieres,Giuliani_and_Vignale} over a very broad range of densities, whereas their $1D$ analogue is described by the Luttinger-liquid model~\cite{haldane_jphysc}. Thus inhomogeneous $1D$ fermionic systems appear as an interesting example in which it is appropriate to change the reference system to one that possesses ground-state Luttinger-liquid rather than Fermi-liquid-type correlations~\cite{soft,lima_2003,burke_2004}. 

Several other examples have been discussed in the literature in which either the LDA reference system is not an EL or the auxiliary system of the Kohn-Sham mapping is not an assembly of noninteracting particles. Kohn and co-workers have introduced the concept of the ``edge electron gas"~\cite{kohn_1998} to study electronic edge regions where the single-particle wavefunctions evolve from oscillatory to evanescent. In the presence of broken gauge symmetry, such as in the superconducting state, an appropriate reference system is the EL in the presence of an external pairing field inducing superconducting correlations~\cite{ogk_1988,floris_2005}. A related approach has also been proposed for Bose-Einstein-condensed systems~\cite{griffin_1995}. Similar in spirit to the above mentioned work on DFT for the Hubbard model, is DFT for the Heisenberg model, in which the reference system is a lattice of spins on equivalent sites~\cite{hemo}. Finally, we mention spectral-DFT~\cite{kotliar_2004}, in which the Kohn-Sham noninteracting system is replaced by a suitably chosen interacting system not handled {\it via} the usual LDA. Such departures from the standard EL paradigm substantially expand the range of usefulness of DFT in condensed-matter physics, but also demand the construction and investigation of new classes of functionals. The present work is concerned with the testing of LDA-type density functionals for strongly correlated $1D$ ultracold Fermi gases confined inside an optical lattice. 

From the experimental point of view such systems, 
which are highly tunable and ideally clean, are attracting a great deal of interdisciplinary interest because they allow to realize strongly interacting many-body systems through the manipulation of relevant degrees of freedom other than the bare atom-atom interaction, such as the well depth of the optical lattice that allows to tune the relative strength of hopping to on-site repulsion/attraction~\cite{cirac_2003}. 
In particular the study of these systems may help us understand a number of phenomena that have been predicted in solid-state and condensed-matter physics. Several effects, known in these subfields of physics for decades, 
have already been observed and quantitatively analyzed in ultracold atomic gases trapped in optical lattices. 
Two beautiful examples are the Bloch oscillations under an applied force 
in a $1D$ optical lattice~\cite{bloch_oscillations} and the superfluid-to-Mott insulator transition of a Bose-Einstein condensate in a $3D$ optical lattice~\cite{jaksch_1998,markus_2002}. Typical $1D$ quantum phenomena have also 
been observed in both Bose and Fermi gases. For instance, in the work of Paredes {\it et al.}~\cite{paredes_2004} and of Kinoshita {\it et al.}~\cite{kinoshita_2004} a $^{87}{\rm Rb}$ gas has been used to realize experimentally a Tonks-Girardeau system~\cite{TG_gas}. The more recent preparation of two-component $^{40}{\rm K}$ Fermi gases in a quasi-$1D$ geometry~\cite{moritz_2005} provides a unique possibility to experimentally study phenomena that were predicted a long time ago for electrons in a $1D$ solid-state environment, such as spin-charge separation in Luttinger liquids~\cite{haldane_jphysc,Giuliani_and_Vignale} and charge-density waves in Luther-Emery liquids~\cite{luther_emery}. 

In this work we focus our attention on a particular $1D$ lattice system of ultracold atoms: a two-component Fermi gas with repulsive intercomponent interactions in the presence of static external potentials that break the lattice translational invariance. Theoretical studies of this model have 
been carried out both by numerical techniques~\cite{rigol_prl,rigol_pra,machida_2004} 
and by LDA-based calculations~\cite{xia_ji_2005,gao_2005,vivaldo_klaus_2005} 
that will be discussed in detail later in this work. 
Building upon the earlier ideas described in Refs.~\onlinecite{soft,lima_2003}, we here 
employ a lattice DFT scheme in which the xc potential is determined exactly at the LDA level through the Bethe-Ansatz solution of the homogeneous $1D$ Hubbard model. The results are tested against accurate QMC simulation data over a broad range of values for the Hubbard on-site interaction, the number of atoms and lattice sites, and different types of external potential. 

The contents of the paper are briefly described as follows. In Sect.~\ref{model} we introduce the lattice Hamiltonian that we use to describe the system of physical interest. For the benefit of readers who are not familiar with the Bethe-Ansatz and the Luttinger liquid, we also briefly summarize the properties of the model and its solution in the absence of external potentials. In Sect.~\ref{soft} we present the self-consistent lattice DFT scheme that we use to deal with the inhomogeneous system under confinement and explain in detail the Bethe-Ansatz LDA that we employ for the xc potential. In Sect.~\ref{numerical_results} we report and discuss our main theoretical results in comparison with QMC simulation data. Finally, in Sect.~\ref{discussion_conclusions} we summarize our main conclusions. An Appendix 
contains the formal derivation of the lattice Kohn-Sham equations.

\section{The fermionic Hubbard model}
\label{model}

We consider a two-component interacting Fermi gas with $N_{\rm f}$ particles which are constrained to move 
under confinement inside a $1D$ lattice with unit lattice constant and $N_{\rm s}$ lattice points labeled by the discrete coordinate $z_i=i, i\in[1,N_{\rm s}]$. This system is described by the following single-band Hubbard Hamiltonian~\cite{jaksch_1998,hofstetter_2002},
\begin{widetext}
\begin{eqnarray}\label{eq:hubbard}
{\hat {\cal H}}&=&-\sum_{i,j}\sum_\sigma
t_{i,j}\left[{\hat c}^{\dagger}_{\sigma}(z_i){\hat c}_{\sigma}(z_j)+{\rm H}.{\rm c}.\right]
+U\sum_{i}\,{\hat n}_{\uparrow}(z_i){\hat n}_{\downarrow}(z_i)
+\sum_{i}V_{\rm ext}(z_i){\hat n}(z_i)={\hat {\cal T}}+{\hat {\cal H}}_{\rm int}+{\hat {\cal H}}_{\rm ext}\,,
\end{eqnarray}
\end{widetext}
where $t_{i,j}=t>0$ if $i$ and $j$ are nearest-neighbor sites and zero otherwise, and $\sigma=\uparrow,\downarrow$ represents a pseudospin-$1/2$ degree of freedom (hyperfine-state label). The field operator ${\hat c}^{\dagger}_{\sigma}(z_i)$ (${\hat c}_{\sigma}(z_i)$) creates (destroys) a fermion with pseudospin $\sigma$ at position $z_i$, ${\hat n}_\sigma(z_i)={\hat c}^{\dagger}_{\sigma}(z_i){\hat c}_{\sigma}(z_i)$ is the pseudospin-resolved site occupation operator normalized to the number of particles with pseudospin $\sigma$, $N_\sigma=\langle \sum_i {\hat n}_\sigma(z_i)\rangle$, and ${\hat n}(z_i)=\sum_\sigma {\hat n}_\sigma(z_i)$ is the total site occupation operator with $\langle \sum_i{\hat n}(z_i)\rangle=N_{\rm f}$. Finally $V_{\rm ext}(z_i)$ is an external static potential associated with the confinement. The Hubbard Hamiltonian without confinement has, of course, 
been widely used in studies of strongly correlated electrons, where the site index
refers to ion positions. It also applies to an electron liquid in a quantum wire under the effect of 
a spatially modulated electric potential~\cite{mansour_2004}, where the site index refers to minima of the superlattice structure. Parabolic confinement is easily achieved in nanowire quantum dots~\cite{samuelson_2003}, and most of the results reported below also apply to that case.

The physical processes associated with each term in Eq.~(\ref{eq:hubbard}) are clear. 
${\hat {\cal T}}$ describes kinetic processes of atom hopping with site-to-site tunneling amplitude $t$. 
${\hat {\cal H}}_{\rm int}$ is the interspecies on-site Hubbard interaction with strength $U$. The intraspecies scattering can be assumed to be negligible because atoms with the same pseudospin are kept apart by the Pauli principle and can thus be taken, to a very good approximation, as noninteracting in an ultracold gaseous state. ${\hat {\cal H}}_{\rm ext}$ gives the coupling of the atoms to a static external potential. In this work we will restrict our attention to repulsive interactions, {\it i.e.} $U>0$, in symmetric systems with equal numbers of atoms of each pseudospin species ($N_\uparrow=N_\downarrow=N_{\rm f}/2$). Attractive interactions have been discussed in Ref.~\onlinecite{gao_2005} and asymmetric Fermi gases in Ref.~\onlinecite{bakhtiari_2005}. 

In the absence of a longitudinal external field ($V_{\rm ext}=0$), ${\hat {\mathcal H}}$ reduces to the Hamiltonian of a $1D$ homogeneous Hubbard model (HHM) that has been solved exactly more than 30 years ago by Lieb and Wu~\cite{lieb_wu}. At zero temperature and for $N_\uparrow=N_\downarrow$ the properties of the $1D$ HHM in the thermodynamic limit ($N_{\rm f},N_{\rm s} \rightarrow \infty$) are determined by two parameters only, the filling factor $n=N_{\rm f}/N_{\rm s}$ and the dimensionless coupling constant $u=U/t$.

\begin{figure}
\begin{center}
\includegraphics[width=1.00\linewidth]{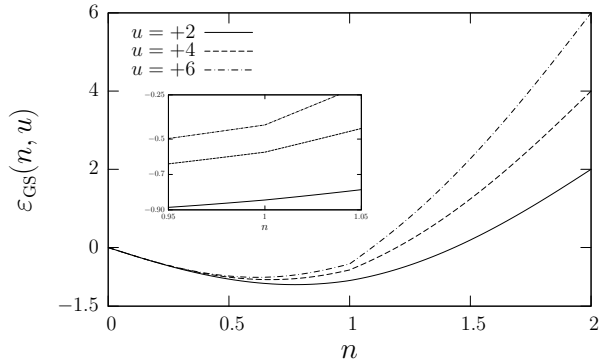}
\caption{Ground-state energy of the repulsive $1D$ homogeneous Hubbard model (in units of hopping parameter $t$) as a function of the filling factor $n$ for various values of the coupling parameter $u$. The cusp at $n=1$ signals the Mott-insulating phase. The inset shows an enlargement of the region $0.95\leq n \leq 1.05$.\label{fig:one}}
\end{center}
\end{figure}

According to Lieb and Wu~\cite{lieb_wu}, 
the ground state (GS) of the repulsive $1D$ HHM in the thermodynamic limit is described by two continuous distribution functions $\rho(x)$ and $\sigma(y)$ which satisfy the Bethe-Ansatz (BA) coupled integral equations,
\begin{widetext}
\begin{equation}\label{eq:lw_1}
\rho(x)=\frac{1}{2\pi}+\frac{\cos{x}}{\pi}\int_{-\infty}^{+\infty}\frac{u/4}{(u/4)^2+(y-\sin{x})^2}\sigma(y)dy
\end{equation}
and
\begin{equation}\label{eq:lw_2}
\sigma(y)=\frac{1}{\pi}\int_{-Q}^{+Q}\frac{u/4}{(u/4)^2+(y-\sin{x})^2}\rho(x)dx-\frac{1}{\pi}
\int_{-\infty}^{+\infty}\frac{u/2}{(u/2)^2+(y-y')^2}\sigma(y')dy'\,.
\end{equation}
\end{widetext}
The parameter $Q$ is determined by the normalization condition 
$\int_{-Q}^{+Q}\rho(x)dx=n$, while $\sigma(y)$ is normalized according to 
$\int_{-\infty}^{+\infty}\sigma(y)dy=n/2$. 
The GS energy of the system (per site) is given by
\begin{equation}\label{eq:gsenergy}
\varepsilon_{\rm \scriptscriptstyle GS}(n\leq1,u)=-2 t\,\int_{-Q}^{+Q}\rho(x)\cos{x}\,dx\,.
\end{equation}
For repulsive interactions the $1D$ HHM describes a Luttinger liquid~\cite{schulz_1990} if $n\neq 1$ or $2$. 
At half-filling, {\it i.e.} $n=1$, the GS is a Mott insulator for every $u\neq 0$, while for $n=2$ it is a band insulator. The two metallic GS branches for $n<1$ and $n>1$ are connected by particle-hole symmetry, 
$\varepsilon_{\rm \scriptscriptstyle GS}(n>1,u)=(n-1)U+\varepsilon_{\rm \scriptscriptstyle GS}(2-n,u)$. 
Note that the presence of a Mott-insulating GS at half-filling is signalled 
by a cusp in the GS energy at $n=1$, induced by the linear term in 
this relation (see Fig.~\ref{fig:one}). 
Correspondingly the charge excitation spectrum possesses a gap. 

The GS energy is analytically known for $u=0$ (noninteracting fermions) as 
$\varepsilon_{\rm \scriptscriptstyle GS}(n\leq 1,u=0)=-4t\sin{(n\pi/2)}/\pi$, for $u=+\infty$ as 
$\varepsilon_{\rm \scriptscriptstyle GS}(n\leq 1,u\rightarrow +\infty)=-2t\sin{(n\pi)}/\pi$, 
and for every positive value of $u$ at half-filling~\cite{lieb_wu} as
\begin{equation}\label{eq:half_filling}
\varepsilon_{\rm \scriptscriptstyle GS}(n=1,u)=-4t\int_0^{+\infty}\frac{{\rm J}_0(x){\rm J}_1(x)}{x[1+\exp{(x u/2)}]}dx\,.
\end{equation}
In Fig.~\ref{fig:one} we show $\varepsilon_{\rm \scriptscriptstyle GS}$ as a function of $n$ for various 
values of $u$. The inset in Fig.~\ref{fig:one} shows the cusp at $n=1$.

In the next Section we show how the GS properties of the 
inhomogeneous system described by Eq.~(\ref{eq:hubbard}) can be calculated very accurately in a DFT scheme based on Eqs.~(\ref{eq:lw_1})-(\ref{eq:half_filling}).

\section{Lattice density-functional theory}
\label{soft}

A powerful tool to calculate the GS properties of a Hamiltonian such as given in Eq.~(\ref{eq:hubbard}) is a lattice version of DFT, the so-called site-occupation functional theory (SOFT). This was introduced in pioneering papers by Gunnarsson and Sch\"onhammer~\cite{soft} to study the band-gap problem in the context of {\it ab initio} theories of fundamental energy gaps in semiconductors and insulators~\cite{Giuliani_and_Vignale}. We summarize the key theoretical aspects of SOFT in the Appendix~\ref{appendix_SOFT}, in order for the present paper to be self-contained. 

Within SOFT the exact GS site occupation
$
n_{\rm \scriptscriptstyle GS}(z_i)=\langle{\rm GS}|{\hat n}(z_i)|{\rm GS}\rangle
$
can be obtained by solving self-consistently the lattice Kohn-Sham (KS) equations
\begin{equation}\label{eq:sks}
\sum_{j}[-t_{i,j}+v_{\rm \scriptscriptstyle KS}[n_{\rm \scriptscriptstyle GS}](z_i)\delta_{ij}]\varphi_\alpha(z_j)=\varepsilon_\alpha\varphi_\alpha(z_i)
\end{equation}
with $v_{\rm \scriptscriptstyle KS}[n_{\rm \scriptscriptstyle GS}](z_i)=U n_{\rm \scriptscriptstyle GS}(z_i)/2+v_{\rm xc}(z_i)+V_{\rm ext}(z_i)$, together with the closure
\begin{equation}\label{eq:closure}
n_{\rm \scriptscriptstyle GS}(z_i)=\sum_{\alpha, {\rm occ.}}\Gamma_\alpha\left|\varphi_\alpha(z_i)\right|^2\,.
\end{equation}
Here the sum runs over the occupied orbitals and the degeneracy factors $\Gamma_\alpha$ satisfy the sum rule 
$\sum_\alpha \Gamma_\alpha=N_{\rm f}$. The first term in the effective Kohn-Sham potential $v_{\rm \scriptscriptstyle KS}$ is the Hartree mean-field contribution, while $v_{\rm xc}[n_{\rm \scriptscriptstyle GS}](z_i)=\delta {\cal E}_{\rm xc}[n]/\delta n(z_i)|_{\rm \scriptscriptstyle GS}$ is the xc potential defined by the derivative of the xc energy ${\cal E}_{\rm xc}[n]$ evaluated at the GS site occupation [see Eq.~(\ref{eq:formalvxc})]. Notice that exchange interactions between parallel-pseudospin atoms have been effectively eliminated in the Hubbard model~(\ref{eq:hubbard}) by restricting the model to one orbital per site. Hence parallel-pseudospin interactions are not treated dynamically in solving the Hamiltonian, but are accounted for implicitly {\it via} a restriction of the Hilbert space. To stress the analogy of the present work with {\it ab initio} applications of standard DFT, we nevertheless continue to call ${\cal E}_{\rm xc}[n]$ and $v_{\rm xc}[n_{\rm \scriptscriptstyle GS}]$ the exchange-correlation energy and the exchange-correlation potential, 
but it is understood that the exchange contribution to these quantities is exactly zero. 
The total GS energy of the system is given by
\begin{eqnarray}\label{eq:gs_energy}
{\cal E}_{\rm \scriptscriptstyle GS}[n_{\rm \scriptscriptstyle GS}]&=&\sum_\alpha\Gamma_\alpha\varepsilon_\alpha-\sum_i v_{\rm xc}(z_i)n_{\rm \scriptscriptstyle GS}(z_i)\nonumber\\
&-&\sum_i U n^2_{\rm \scriptscriptstyle GS}(z_i)/4+{\cal E}_{\rm xc}[n_{\rm \scriptscriptstyle GS}]\,.
\end{eqnarray} 

Equations~(\ref{eq:sks})-(\ref{eq:gs_energy}) provide a formally exact scheme to calculate $n_{\rm \scriptscriptstyle GS}(z_i)$ and ${\cal E}_{\rm \scriptscriptstyle GS}$, but ${\cal E}_{\rm xc}$ needs to be approximated. 
The LDA has been shown to provide an excellent account of the GS properties of a large variety of inhomogeneous systems~\cite{wk,d&g,joulbert_1998,Giuliani_and_Vignale,hk,ks}. In this work we employ the following BA-based LDA (BALDA) functional
\begin{equation}\label{eq:balda}
v^{\rm \scriptscriptstyle BALDA}_{\rm xc}[n_{\rm \scriptscriptstyle GS}](z_i)=\left. v^{\rm hom}_{\rm xc}(n,u)\right|_{n\rightarrow n_{\rm \scriptscriptstyle GS}(z_i)}\,,
\end{equation}
where, in analogy with {\it ab initio} DFT, the xc potential $v^{\rm hom}_{\rm xc}(n,u)$ of the $1D$ HHM is defined by
\begin{equation}\label{eq:vxchom}
v^{\rm hom}_{\rm xc}(n,u)=\frac{\partial}{\partial n}\left[\varepsilon_{\rm \scriptscriptstyle GS}(n,u)-\varepsilon_{\rm \scriptscriptstyle GS}(n,0)-\frac{U}{4}n^2\right]\,.
\end{equation}
Thus, within the LDA scheme proposed in Eqs.~(\ref{eq:balda}) and~(\ref{eq:vxchom}), the only necessary input is the xc potential of the $1D$ HHM, which is known from its BA solution. 

\subsection{The exchange-correlation potential of the $1D$ HHM}

In what follows we propose two alternative ways to calculate the xc potential of the HHM. 

\subsubsection{${\rm BALDA}/{\rm LSOC}$}

A semi-analytical scheme, in which the calculation of $v^{\rm hom}_{\rm xc}(n,u)$ is carried out with an accurate parametrization formula for $\varepsilon_{\rm \scriptscriptstyle GS}(n,u)$, has been proposed by Lima {\it et al.} (LSOC)~\cite{lima_2003}. This is very similar in spirit to what is done in the EL-based LDA calculations on $3D$ and $2D$ electronic systems~\cite{gatto_2005}, where the only input is the xc energy of the EL for which accurate parametrizations are available. Results for $n_{\rm \scriptscriptstyle GS}(z_i)$ that are obtained 
with $v^{\rm hom}_{\rm xc}(n,u)$ determined according to this 
semi-analytical route will be labeled by the acronym ${\rm BALDA}/{\rm LSOC}$.

\subsubsection{${\rm BALDA}/{\rm FN}$}
\label{baldasubsubsec}

\begin{figure}
\begin{center}
\includegraphics[width=1.00\linewidth]{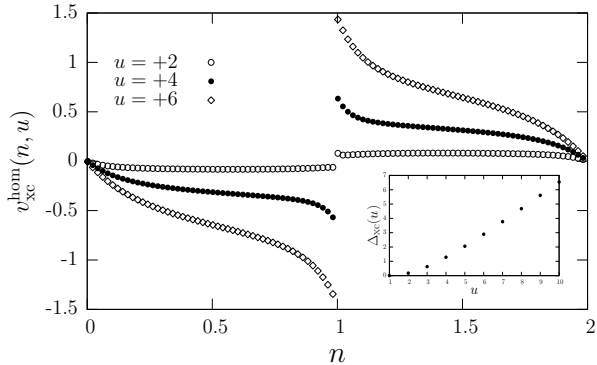}
\caption{The exchange-correlation potential of the repulsive $1D$ homogeneous Hubbard model 
(in units of $t$) as a function of $n$ for various values of $u$. The inset shows the discontinuity $\Delta_{\rm xc}(u)$ (in units of $t$) as a function of $u$.\label{fig:two}}
\end{center}
\end{figure}

A very appealing feature of Eqs.~(\ref{eq:balda}) and~(\ref{eq:vxchom}) from the formal DFT viewpoint, is that one can go a step further than the usual parametrized LDA and establish a fully numerical improvement over the ${\rm BALDA}/{\rm LSOC}$ scheme. This procedure does not rely on any approximation for $v^{\rm hom}_{\rm xc}(n,u)$ and can easily be set up by observing that $v^{\rm hom}_{\rm xc}(n,u)$ satisfies the exact BA equation
\begin{eqnarray}\label{eq:balda_exact}
v^{\rm hom}_{\rm xc}(n<1,u)&=&-4 t\int_{0}^{+Q}[\partial_n\rho(x)]
\cos{x}\,dx\nonumber\\
&-&4t\rho(Q)(\partial_n Q)\cos{Q}+\delta v_{\rm KH}
\end{eqnarray}
and the symmetry $v^{\rm hom}_{\rm xc}(n>1,u)=-v^{\rm hom}_{\rm xc}(2-n,u)$. In Eq.~(\ref{eq:balda_exact}) we have $\delta v_{\rm KH}=2t \cos{(n\pi/2)}-Un/2$. Equation~(\ref{eq:balda_exact}) must be supplemented by a set of exact BA equations for $\partial_n\rho$, $\partial_n\sigma$, and $\partial_n Q$, which can be derived from Eqs.~(\ref{eq:lw_1})-(\ref{eq:lw_2}) upon differentiating with respect to $n$~\cite{kocharian_1999}. Illustrative numerical results for $v^{\rm hom}_{\rm xc}(n,u)$ are reported in Fig.~\ref{fig:two}. 
Results for $n_{\rm \scriptscriptstyle GS}(z_i)$ that are obtained 
with $v^{\rm hom}_{\rm xc}(n,u)$ determined according to this fully numerical route will be labeled by 
the acronym ${\rm BALDA}/{\rm FN}$. 

A second appealing feature of the local scheme in Eqs.~(\ref{eq:balda}) and~(\ref{eq:vxchom}) is that 
the Mott cusp in $\varepsilon_{\rm \scriptscriptstyle GS}(n,u)$ is responsible for an intrinsic discontinuity $\Delta_{\rm xc}(u)$ in $v^{\rm hom}_{\rm xc}$ at $n=1$,
\begin{eqnarray}\label{eq:delta_xc}
\Delta_{\rm xc}(u)&=&\lim_{n\rightarrow 1^+}v^{\rm hom}_{\rm xc}(n,u)-\lim_{n\rightarrow 1^-}v^{\rm hom}_{\rm xc}(n,u)\nonumber\\
&=&U-2\lim_{n\rightarrow 1^-}\partial_n \varepsilon_{\rm \scriptscriptstyle GS}(n<1,u)
\end{eqnarray}
(see the inset in Fig.~\ref{fig:two}). As a consequence and contrary to the EL-based LDA, the xc potential in Eq.~(\ref{eq:balda}) possesses a discontinuity in its derivative~\cite{derivative_discontinuity,lima_2002,Giuliani_and_Vignale}.

Correlation-induced discontinuities, here related to Mott-transition physics, also appear in the xc potential 
of a $2D$ EL in the fractional quantum Hall regime, where correlation-induced gaps at fractional filling factors are associated with the formation of an incompressible liquid (see {\it e.g.} Fig. 10.28 in Ref.~\cite{Giuliani_and_Vignale}). These physical discontinuities make it difficult to obtain converging self-consistent solutions of the KS equations whenever the local density reaches a value associated with a gap in the homogeneous reference fluid~\cite{ferconi_1995}. 
In our case this happens only when $n_{\rm \scriptscriptstyle GS}(z_i)\rightarrow 1$ at some position $z_i$. 
In the context of DFT calculations of the edge structure of fractional quantum Hall liquids, Ferconi {\it et al.}~\cite{ferconi_1995} have handled this convergence problem by going to a very small finite temperature. This allows fractional occupation of the single-particle KS levels. 

\subsubsection{${\rm TLDA}$}

A conceptually simpler route to solve such convergence problems, which we have examined in this work, is to resort to an LDA also for the noninteracting kinetic energy functional ${\cal T}_{\rm s}[n]$ (see the Appendix~\ref{appendix_SOFT}). This is similar to the Thomas-Fermi (TF) approximation, which involves an LDA for ${\cal T}_{\rm s}[n]$ but ignores the exchange and correlation energy. This in itself is not a fully reliable approach for the present work, which is directed at strongly correlated regimes. However, we would like to exploit one favorable aspect of the TF approach, {\it i.e.} the replacement of self-consistent solutions by a direct minimization of total energy functionals. To achieve this we combine a TF-like LDA for ${\cal T}_{\rm s}[n]$ with an LDA for ${\cal E}_{\rm xc}[n]$. This amounts to approximating all nontrivial terms in the total energy functional by the LDA, and for this reason we refer to this approach as the total-energy LDA (TLDA)~\cite{vivaldo_klaus_2005}. At variance from Ref.~\onlinecite{vivaldo_klaus_2005}, where TLDA was used in conjunction with the LSOC parametrization of the xc energy, we here employ its fully numerical counterpart~\cite{xia_ji_2005}, in which all results pertaining to the homogeneous reference system are obtained from the Lieb-Wu equations. Both formulations of TLDA circumvent a self-consistent solution of the KS equations, at the expense of a less accurate account of the kinetic energy. The TLDA approximation consists in writing
\begin{eqnarray}\label{eq:tlda}
{\cal E}^{\rm \scriptscriptstyle TLDA}_{\rm \scriptscriptstyle GS}[n_{\rm \scriptscriptstyle GS}]&=&\sum_i 
\left.\varepsilon_{\rm \scriptscriptstyle GS}(n,u)
\right|_{n\rightarrow n_{\rm \scriptscriptstyle GS}(z_i)}\nonumber\\
&+&\sum_{i}V_{\rm ext}(z_i)n_{\rm \scriptscriptstyle GS}(z_i)\,.
\end{eqnarray}
Within this approximation the variational equation~(\ref{eq:HK_variational})
can be written as
\begin{equation}\label{eq:tf_scheme}
\left.\partial_n \varepsilon_{\rm \scriptscriptstyle GS}(n,0)\right|_{n\rightarrow n_{\rm \scriptscriptstyle GS}(z_i)}+v_{\rm \scriptscriptstyle KS}[n_{\rm \scriptscriptstyle GS}](z_i)={\rm constant}\,,
\end{equation}
where the constant is fixed by normalization. Results for $n_{\rm \scriptscriptstyle GS}(z_i)$ that are obtained by this route will be labeled by the acronym ${\rm TLDA}$. 

\section{Numerical results}
\label{numerical_results}

\begin{figure*}
\begin{center}
\tabcolsep=0 cm
\begin{tabular}{cc}
\includegraphics[width=0.50\linewidth]{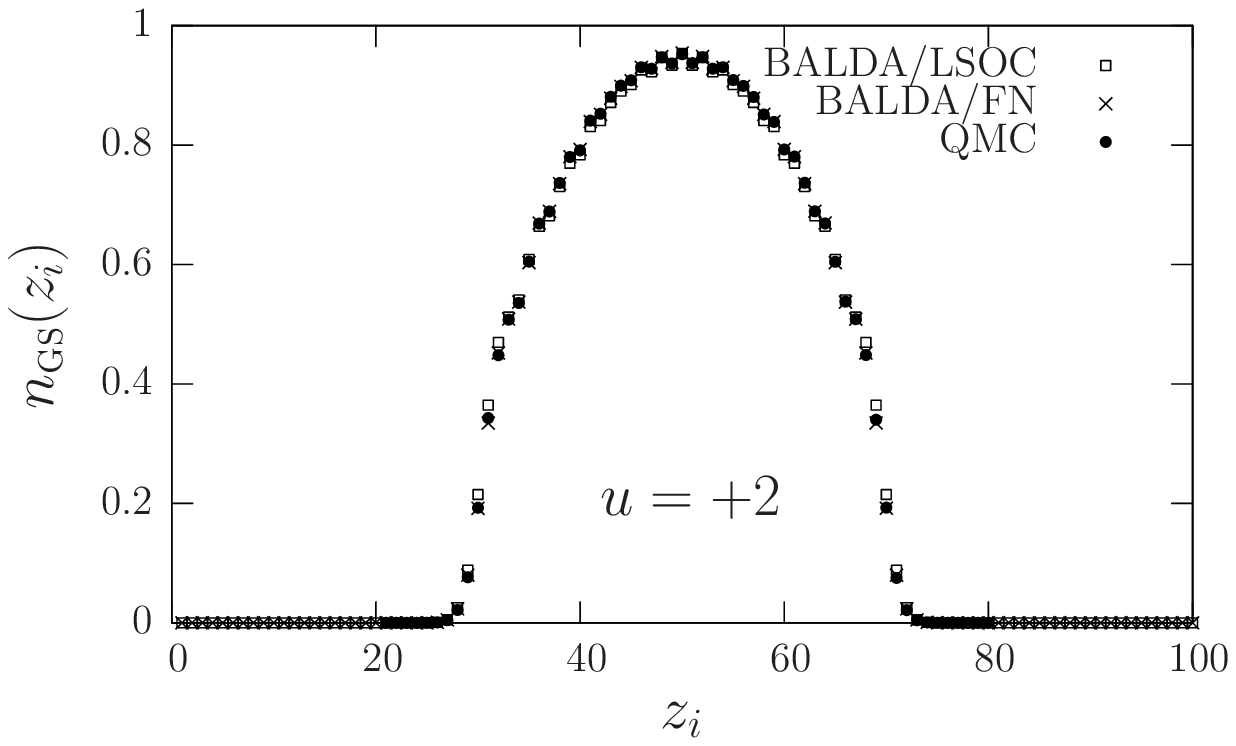}&
\includegraphics[width=0.50\linewidth]{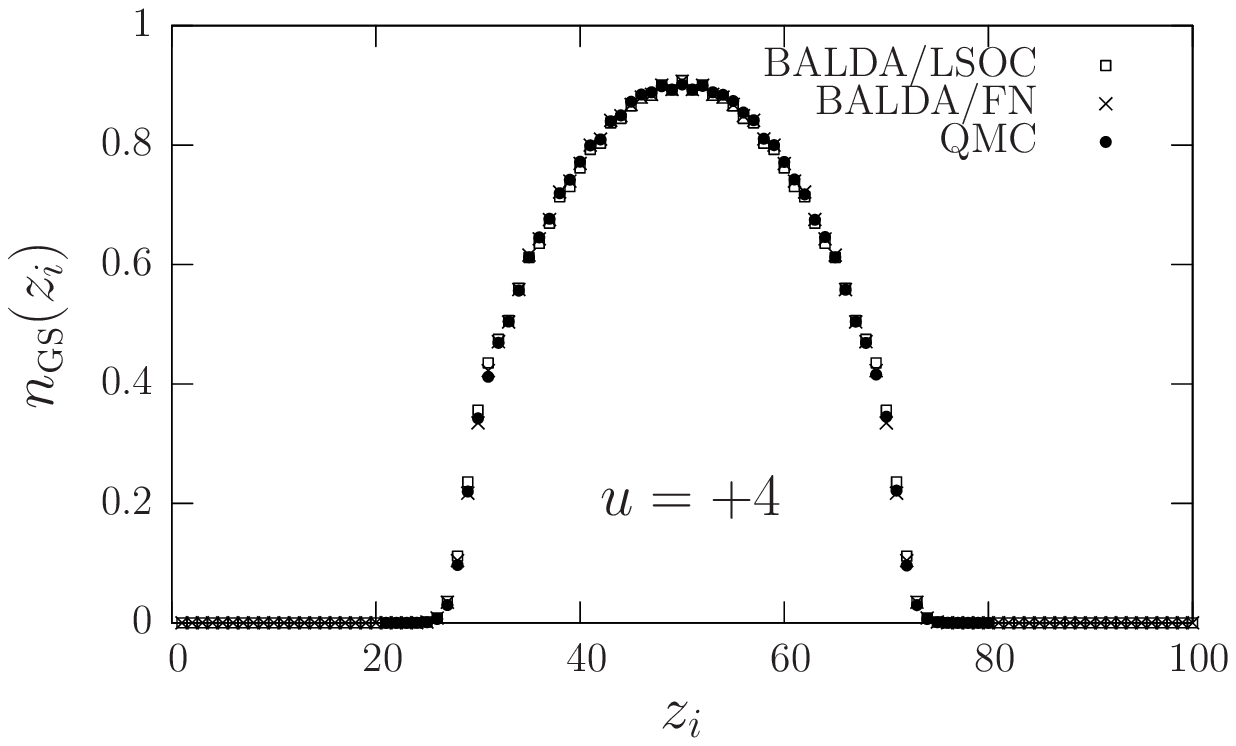}\\
\includegraphics[width=0.50\linewidth]{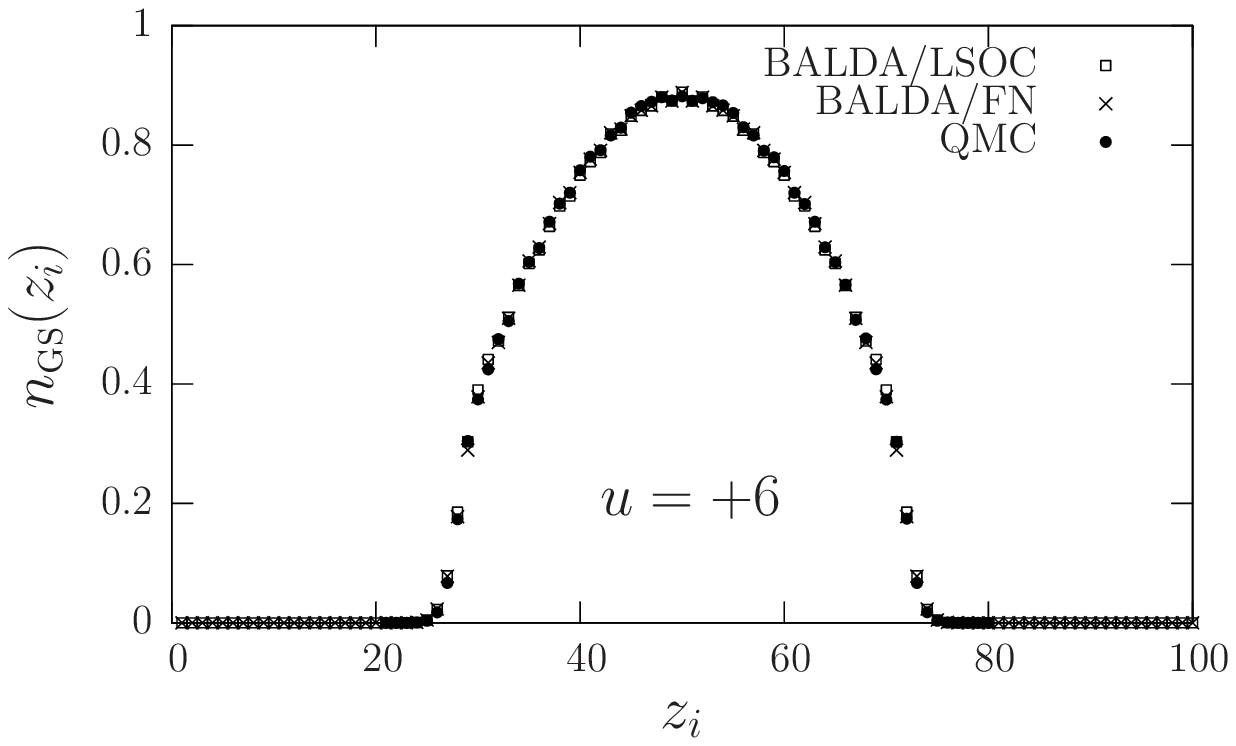}&
\includegraphics[width=0.50\linewidth]{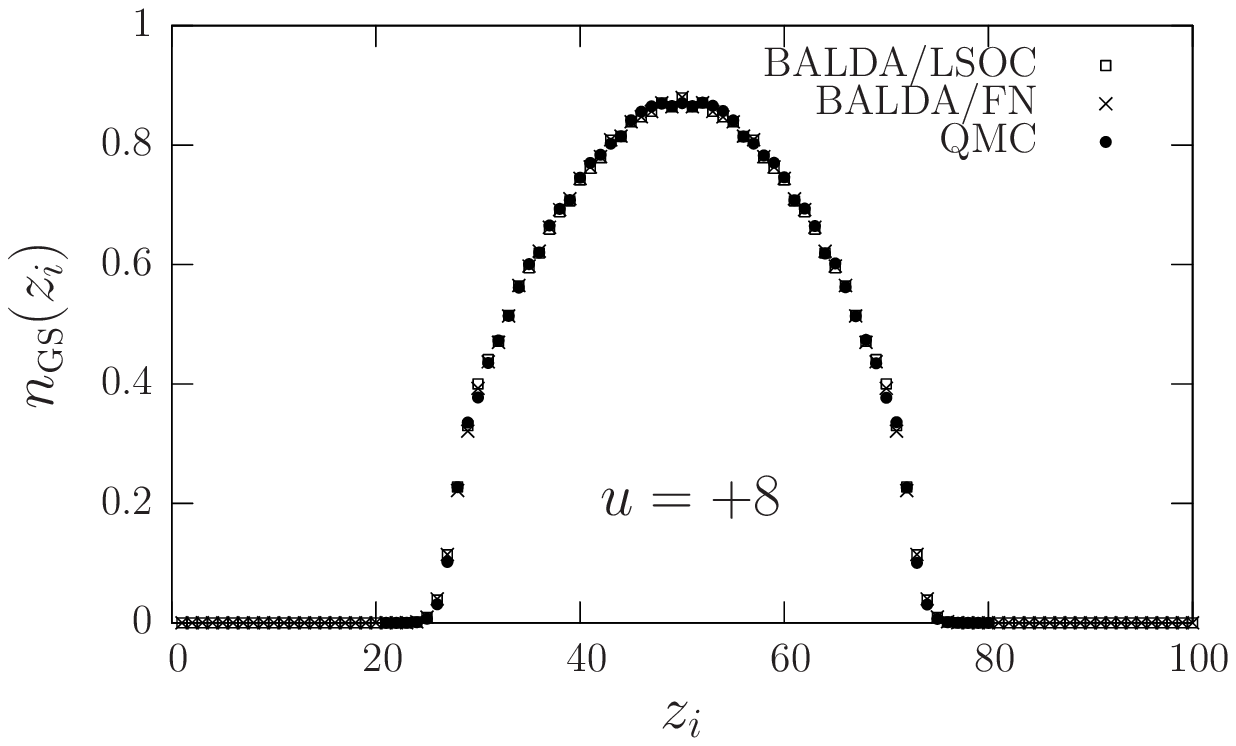}
\end{tabular}
\caption{Site occupation $n_{\rm \scriptscriptstyle GS}(z_i)$ as a function of site position $z_i$ 
for a repulsive Fermi gas with $N_{\rm f}=30$ atoms, trapped in a 
harmonic potential of strength $V_2/t=6\times 10^{-3}$ and in a lattice with $N_{\rm s}=100$ sites. 
The interaction parameter is varied from $u=+2$ in the top left panel to $u=+8$ in the bottom right panel. 
${\rm BALDA}/{\rm LSOC}$ and ${\rm BALDA}/{\rm FN}$ results are compared with the QMC data.\label{fig:three}}
\end{center}
\end{figure*}

\begin{figure}
\begin{center}
\includegraphics[width=1.00\linewidth]{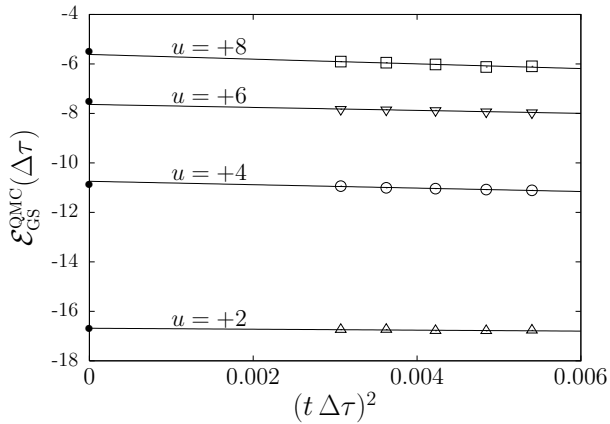}
\caption{QMC ground-state energy (in units of $t$) 
as a function of $(t\,\Delta \tau)^2$, $\Delta\tau$ being the Trotter decomposition paramater, for the same system parameters as in Fig.~\ref{fig:three} (see Appendix~\ref{appendix_QMC}). The QMC error bar at each point is smaller than the size of the symbol. The thin solid lines are a linear fit to the QMC data, which extrapolates to the limit $\Delta\tau\rightarrow 0$. The filled circles at $\Delta\tau=0$ indicate the ${\rm BALDA}/{\rm FN}$ results.\label{fig:four}}
\end{center}
\end{figure}

\begin{figure}
\begin{center}
\includegraphics[width=1.00\linewidth]{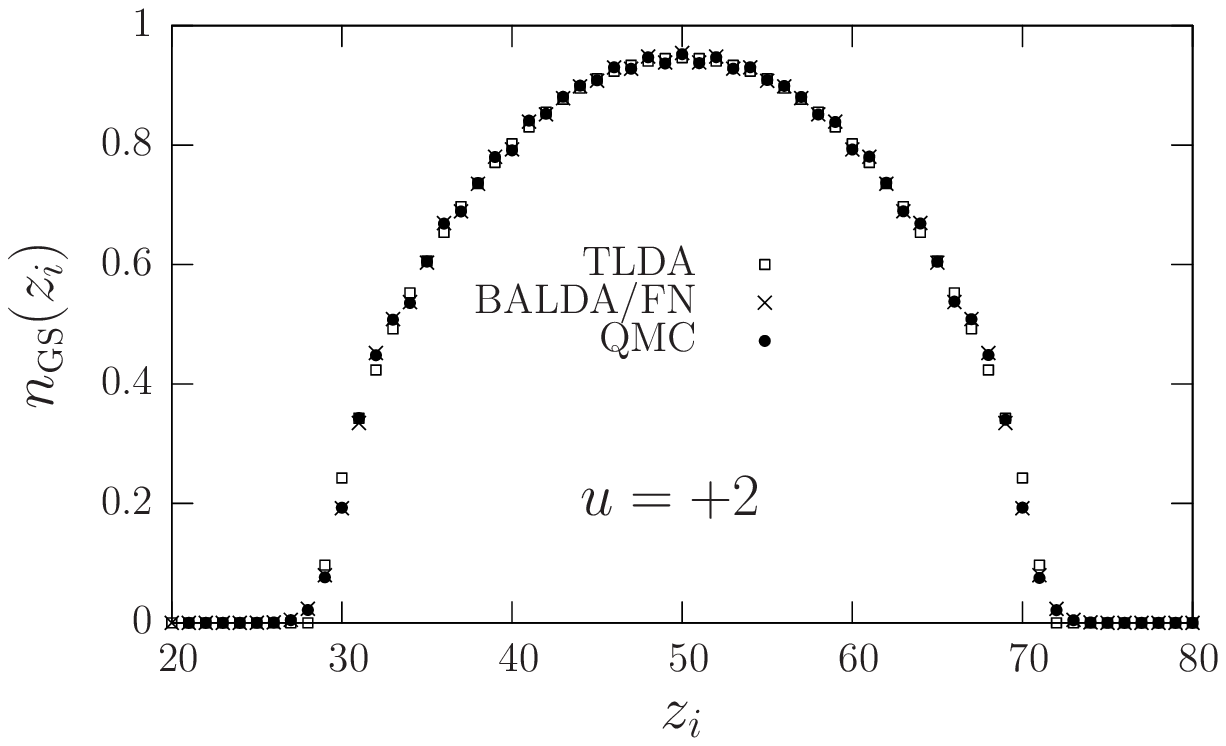}
\includegraphics[width=1.00\linewidth]{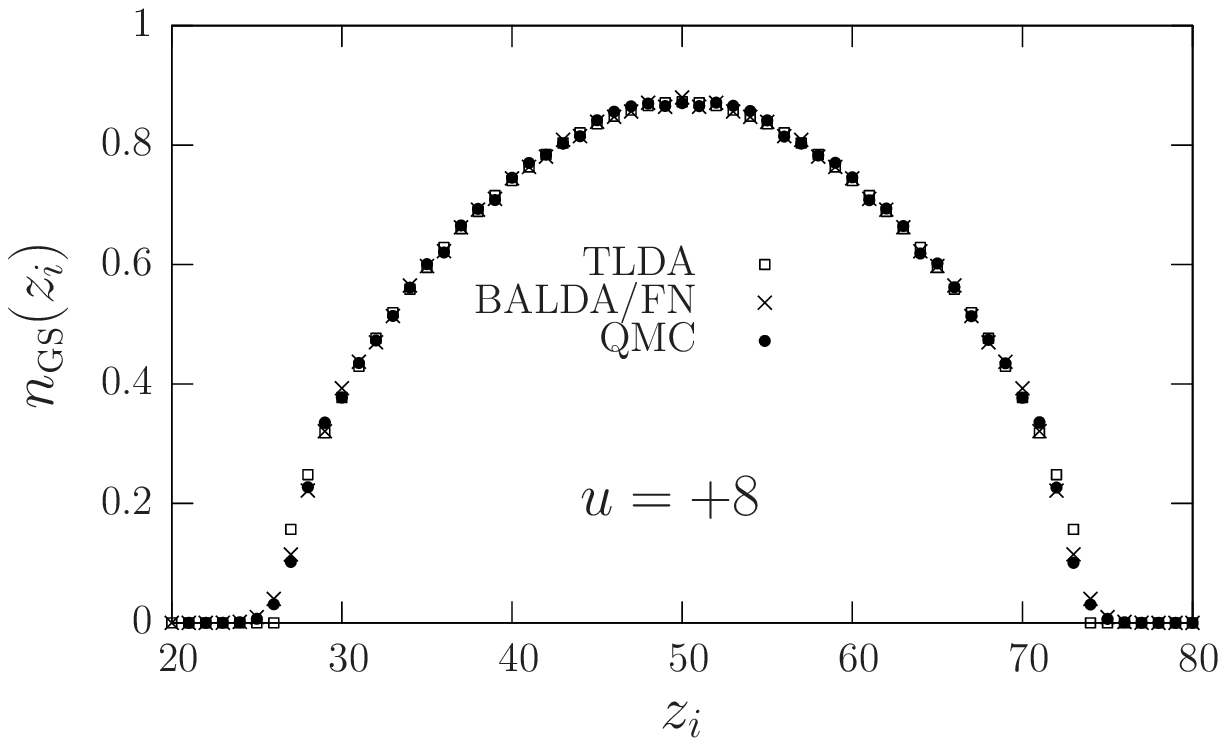}
\end{center}
\caption{Site occupation $n_{\rm \scriptscriptstyle GS}(z_i)$ as a function of $z_i$ 
for a repulsive Fermi gas with $N_{\rm f}=30$ atoms trapped 
in a harmonic potential of strength $V_2/t=6\times 10^{-3}$ and in a lattice with $N_{\rm s}=100$ sites. The interaction parameter is $u=+2$ in the top panel and $u=+8$ in the bottom panel.
${\rm TLDA}$ and ${\rm BALDA}/{\rm FN}$ results are compared with QMC data.\label{fig:five}}
\end{figure}

\begin{figure}
\begin{center}
\includegraphics[width=1.00\linewidth]{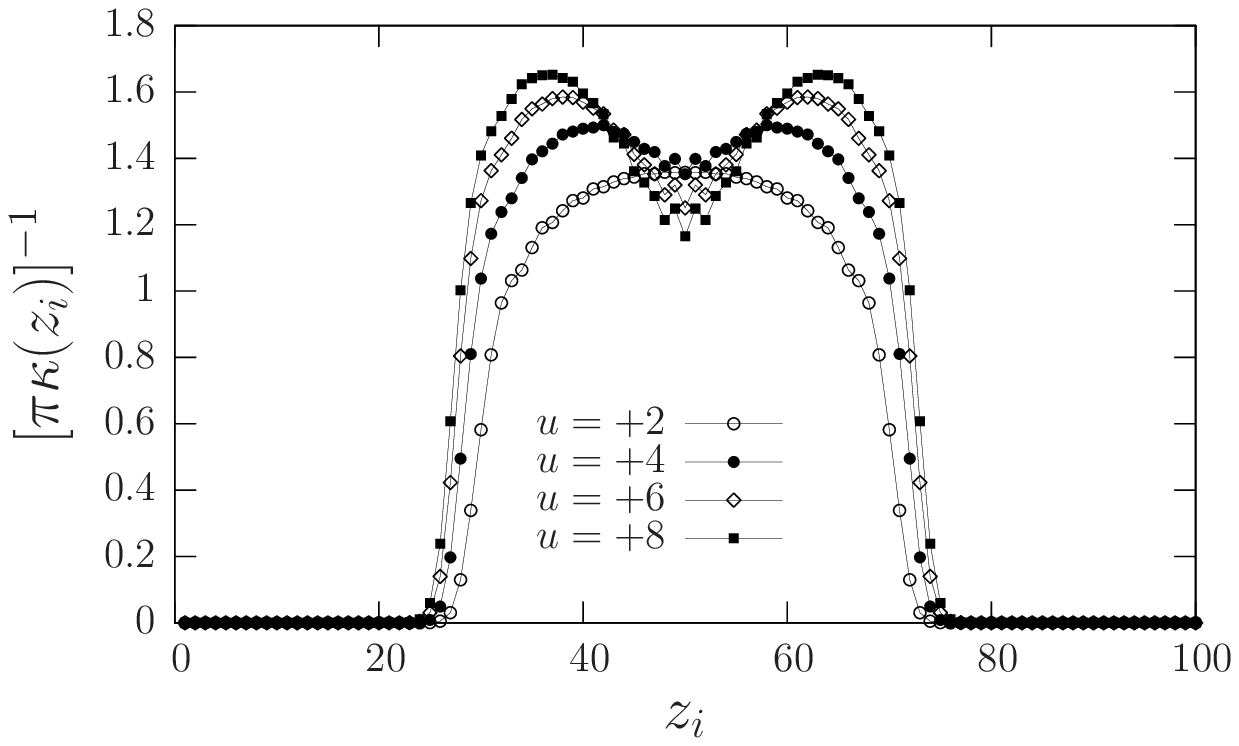}
\caption{${\rm BALDA}/{\rm FN}$ results for the local inverse compressibility 
$[\pi \kappa(z_i)]^{-1}$ (in units of $t$) as a function of $z_i$ 
for the same system parameters as in Fig.~\ref{fig:three}.\label{fig:six}}
\end{center}
\end{figure}

\begin{figure}
\begin{center}
\tabcolsep=0 cm
\begin{tabular}{c}
\includegraphics[width=1.00\linewidth]{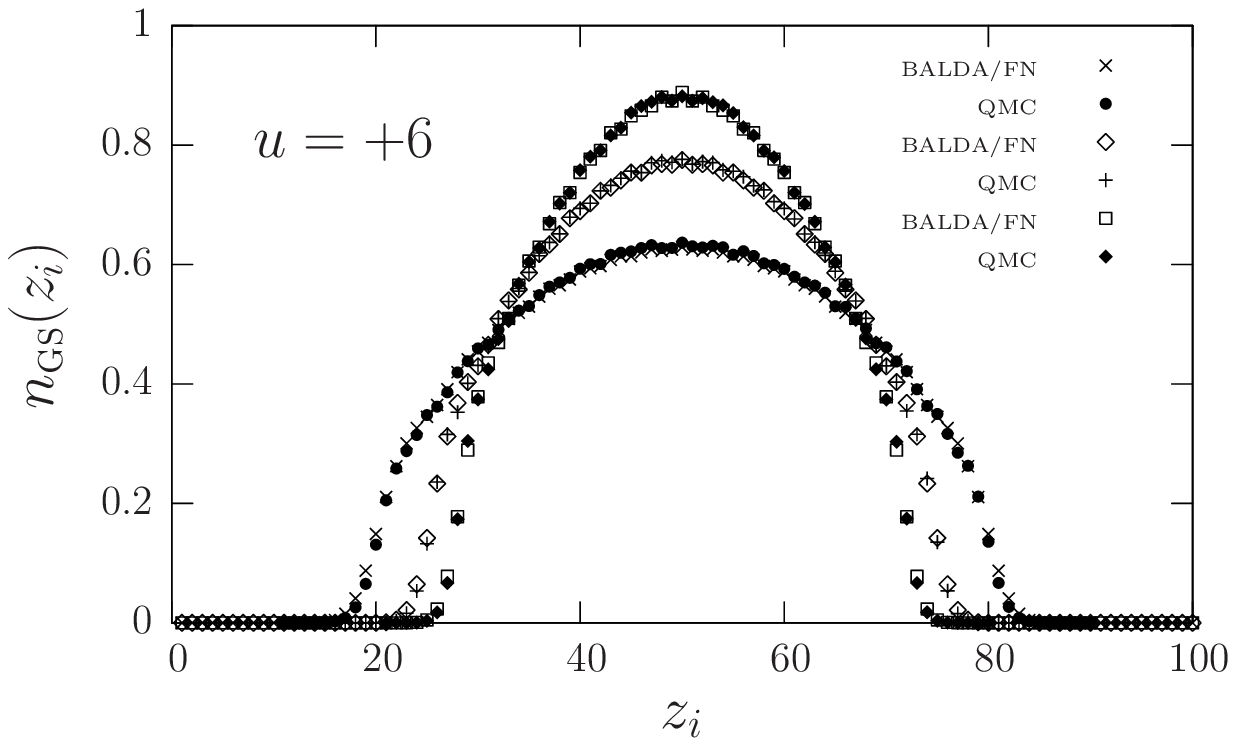}\\
\includegraphics[width=1.00\linewidth]{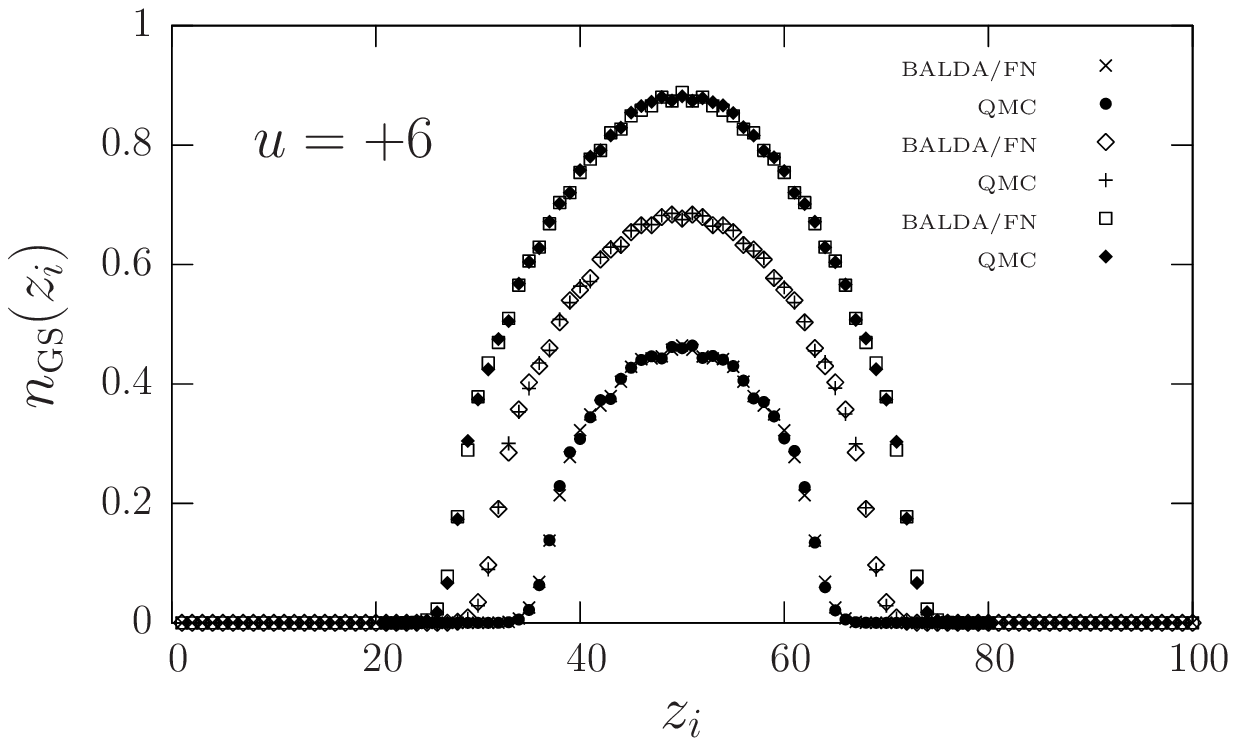}
\end{tabular}
\caption{Top panel: Site occupation $n_{\rm \scriptscriptstyle GS}(z_i)$ as a function of $z_i$ for a repulsive Fermi gas with $N_{\rm f}=30$ atoms, trapped in a harmonic potential with strength 
$V_2/t=2\times 10^{-3},4\times 10^{-3}$, and $6\times 10^{-3}$ (from bottom to top) 
and in a lattice with $N_{\rm s}=100$ sites. 
Bottom panel: Site occupation $n_{\rm \scriptscriptstyle GS}(z_i)$ as a function of $z_i$ for a repulsive Fermi gas with $N_{\rm f}=10,20$, and $30$ atoms (from bottom to top), 
trapped in a harmonic potential of strength $V_2/t=6\times 10^{-3}$ 
and in a lattice with $N_{\rm s}=100$ sites. In both panels results of the 
${\rm BALDA}/{\rm FN}$ scheme at $u=+6$ are compared with QMC data.\label{fig:seven}}
\end{center}
\end{figure}

\begin{figure}
\begin{center}
\tabcolsep=0 cm
\begin{tabular}{c}
\includegraphics[width=1.00\linewidth]{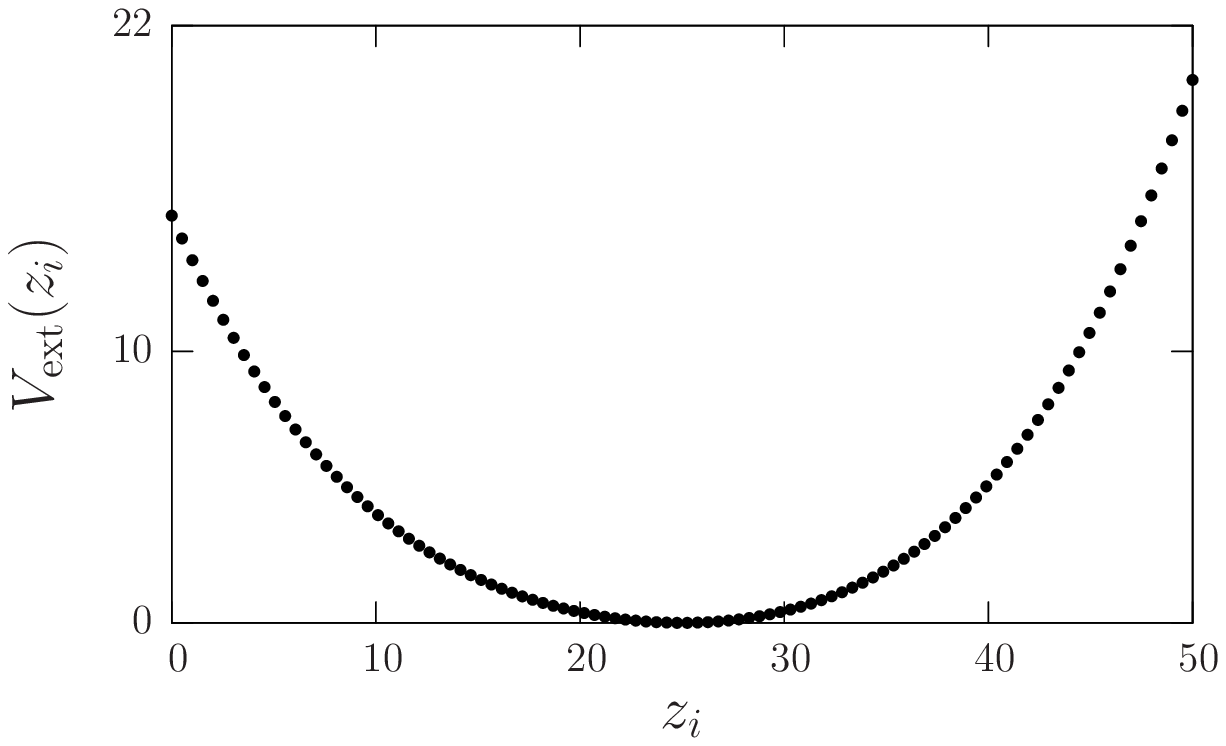}\\
\includegraphics[width=1.00\linewidth]{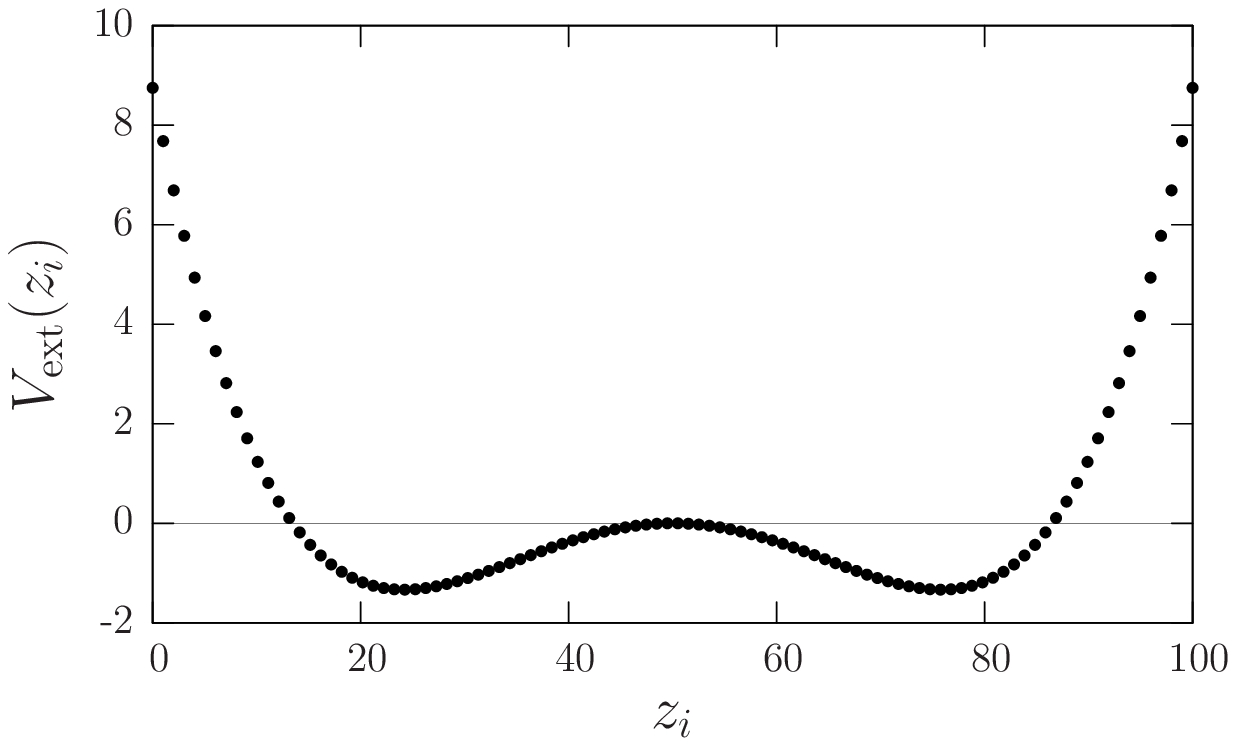}
\end{tabular}
\caption{Top panel: an example of asymmetric trapping potential (in units of $t$) as a function of $z_i$ for the case $V_2/t=1.6\times 10^{-2}$, $V_3/t=1.6\times 10^{-4}$, and $V_4/t=1.92\times 10^{-5}$.
Bottom panel: a double-well potential with $V_2/t=-4\times 10^{-3}$ and $V_4/t=3\times 10^{-6}$.
\label{fig:eight}}
\end{center}
\end{figure}

\begin{figure}
\begin{center}
\tabcolsep=0 cm
\begin{tabular}{c}
\includegraphics[width=1.00\linewidth]{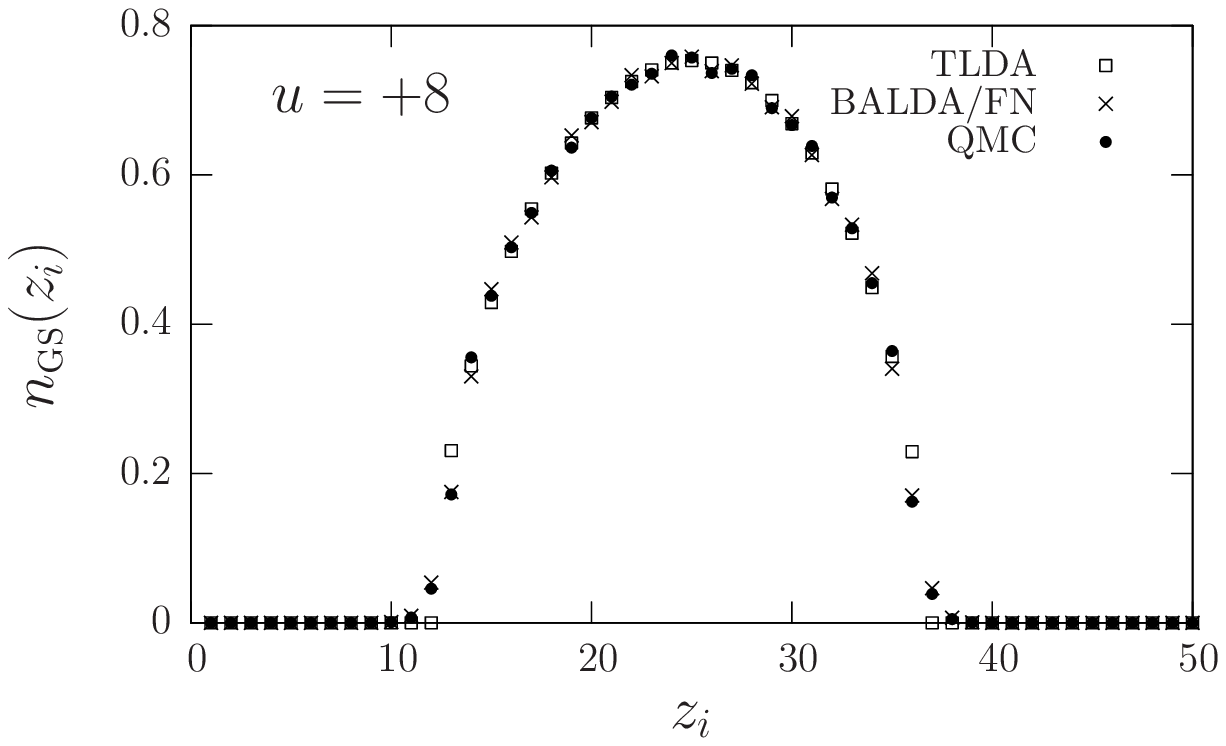}\\
\includegraphics[width=1.00\linewidth]{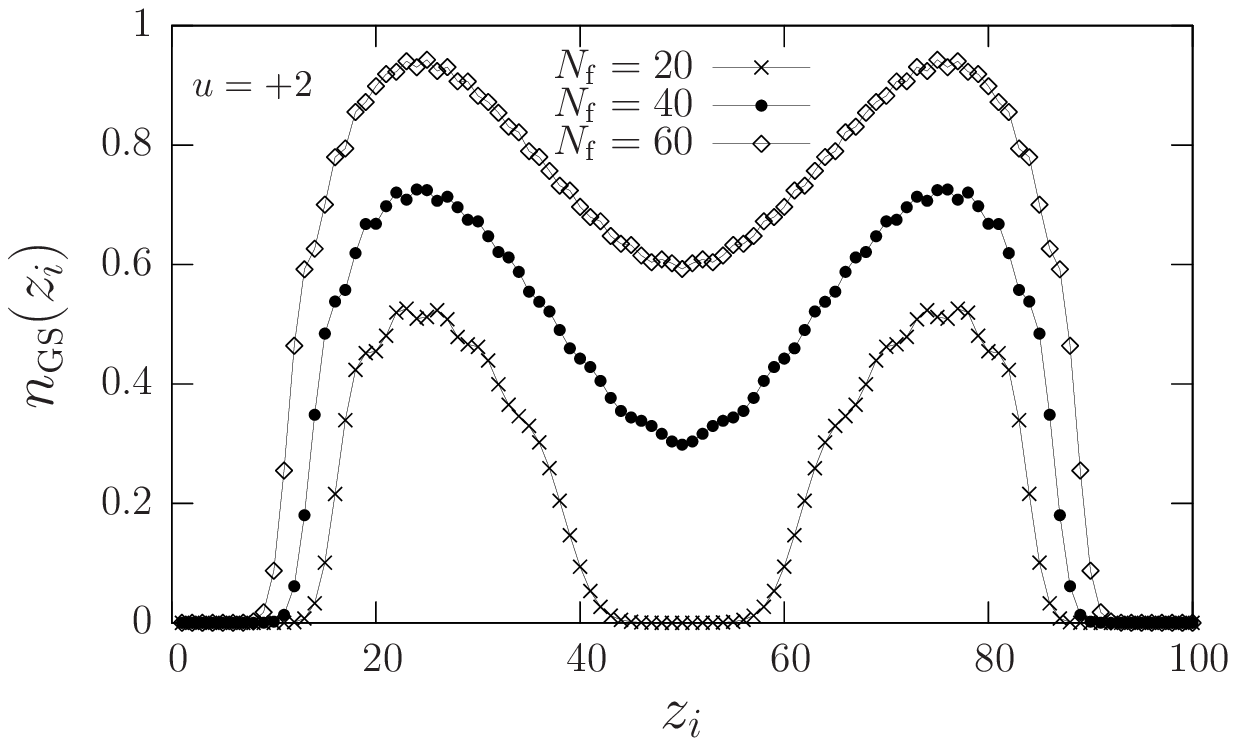}
\end{tabular}
\caption{Top panel: Site occupation $n_{\rm \scriptscriptstyle GS}(z_i)$ as a function of $z_i$ 
for a repulsive Fermi gas with $N_{\rm f}=14$ atoms and $u=+8$, trapped 
in the potential of the top panel of Fig.~\ref{fig:eight} and in a lattice with $N_{\rm s}=50$ sites. 
Results of the TLDA and ${\rm BALDA}/{\rm FN}$ schemes are compared with QMC data.
Bottom panel: Site occupation $n_{\rm \scriptscriptstyle GS}(z_i)$ as a function of $z_i$ 
for a repulsive Fermi gas with $N_{\rm f}=20,40$, and $60$ atoms and $u=+2$, trapped 
in the potential of the bottom panel of Fig.~\ref{fig:eight} and 
in a lattice with $N_{\rm s}=100$ sites. Results of the ${\rm BALDA}/{\rm FN}$ scheme only are presented.\label{fig:nine}}
\end{center}
\end{figure}

\begin{figure}
\begin{center}
\includegraphics[width=1.00\linewidth]{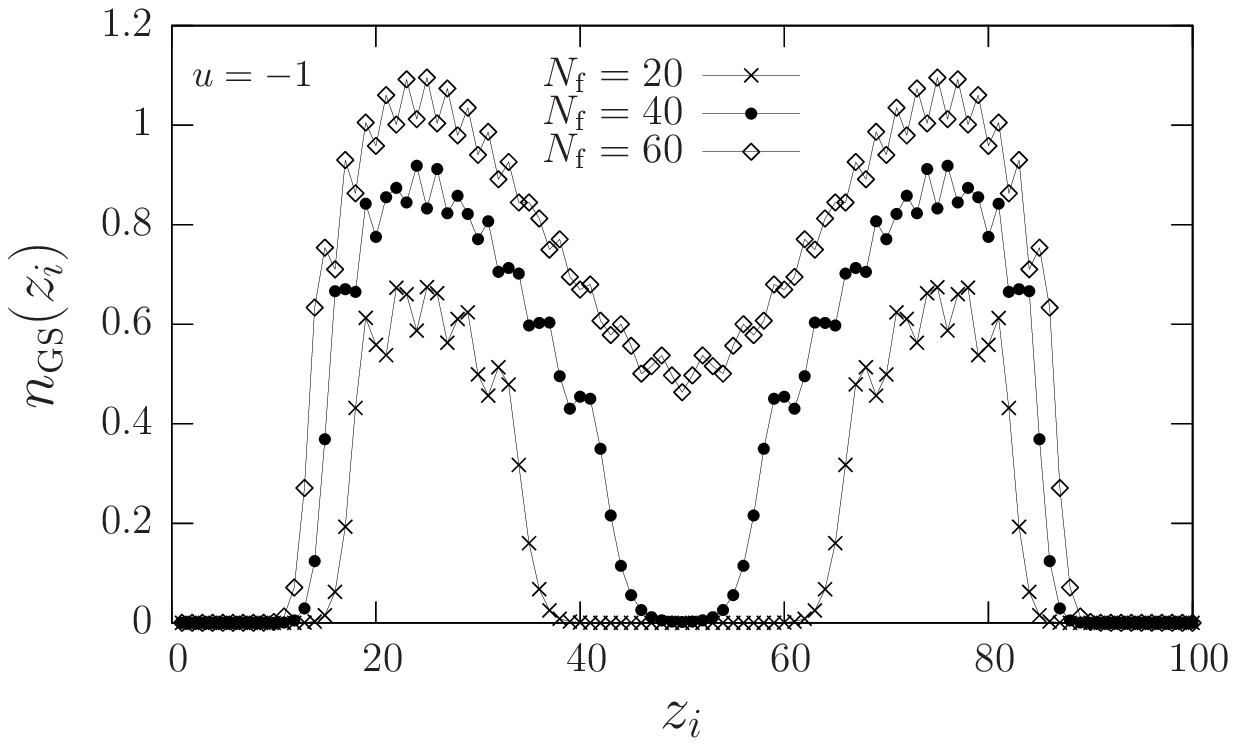}
\caption{Site occupation $n_{\rm \scriptscriptstyle GS}(z_i)$ as a function of $z_i$ 
for an attractive Fermi gas with $N_{\rm f}=20,40$, and $60$ atoms and $u=-1$, trapped 
in the potential of the bottom panel of Fig.~\ref{fig:eight} and in a lattice with $N_{\rm s}=100$ sites.
Results of the ${\rm BALDA}/{\rm FN}$ scheme only are presented.\label{fig:ten}}
\end{center}
\end{figure}

\begin{figure*}
\begin{center}
\tabcolsep=0cm
\begin{tabular}{cc}
\includegraphics[width=0.50\linewidth]{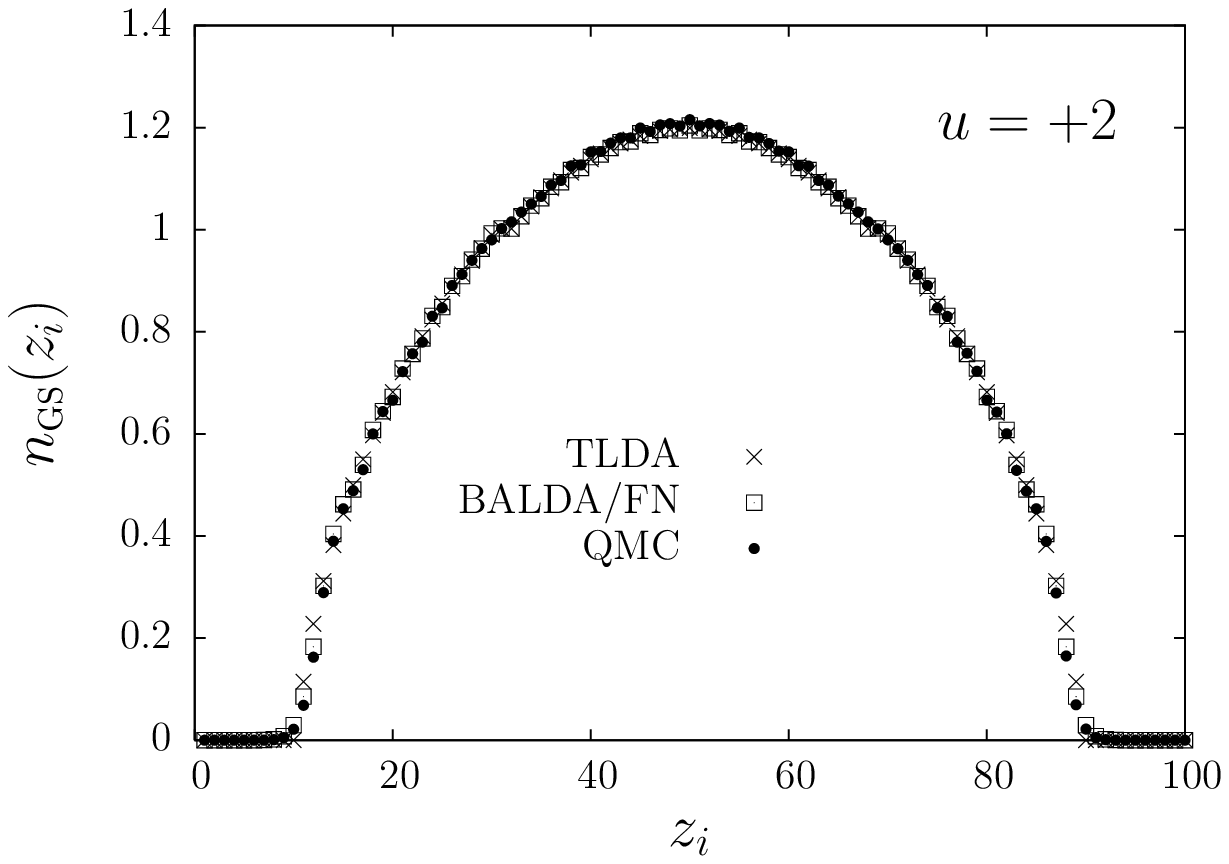}&
\includegraphics[width=0.50\linewidth]{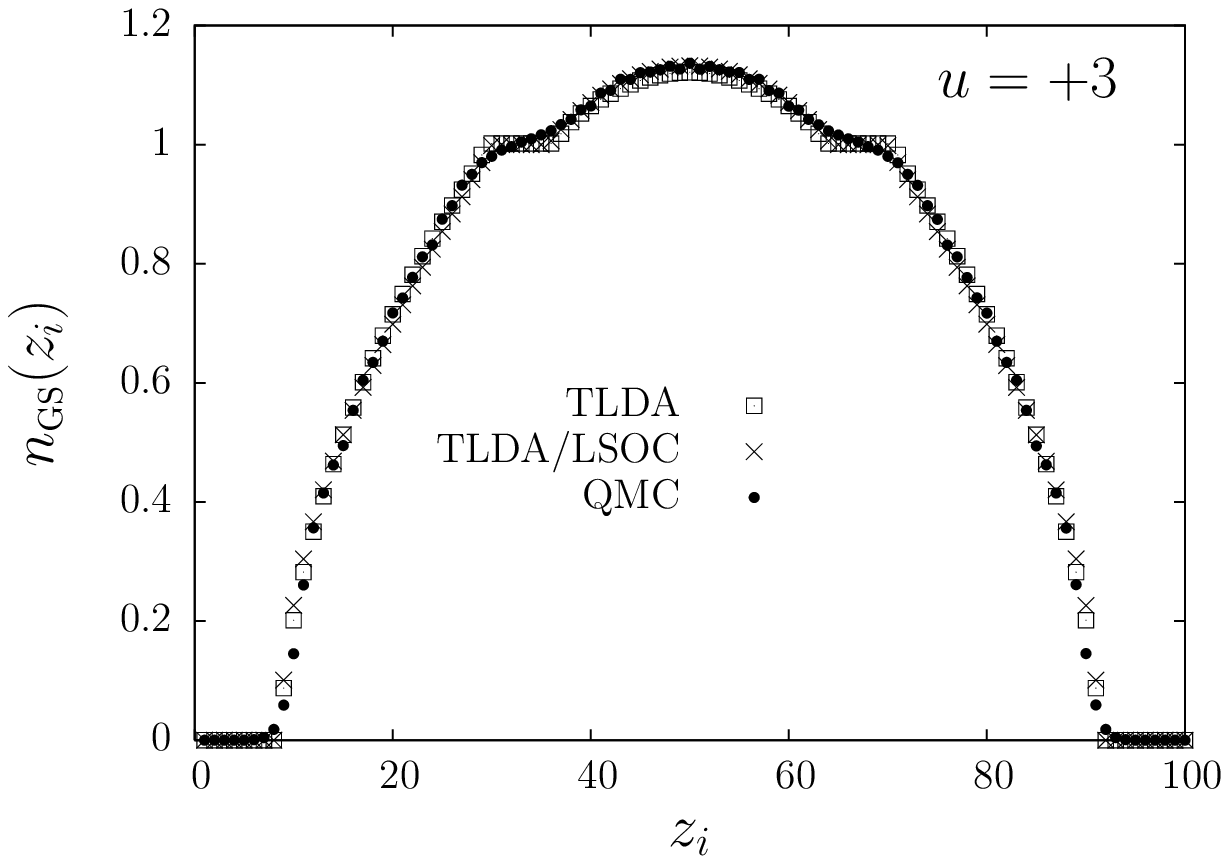}\\
\includegraphics[width=0.50\linewidth]{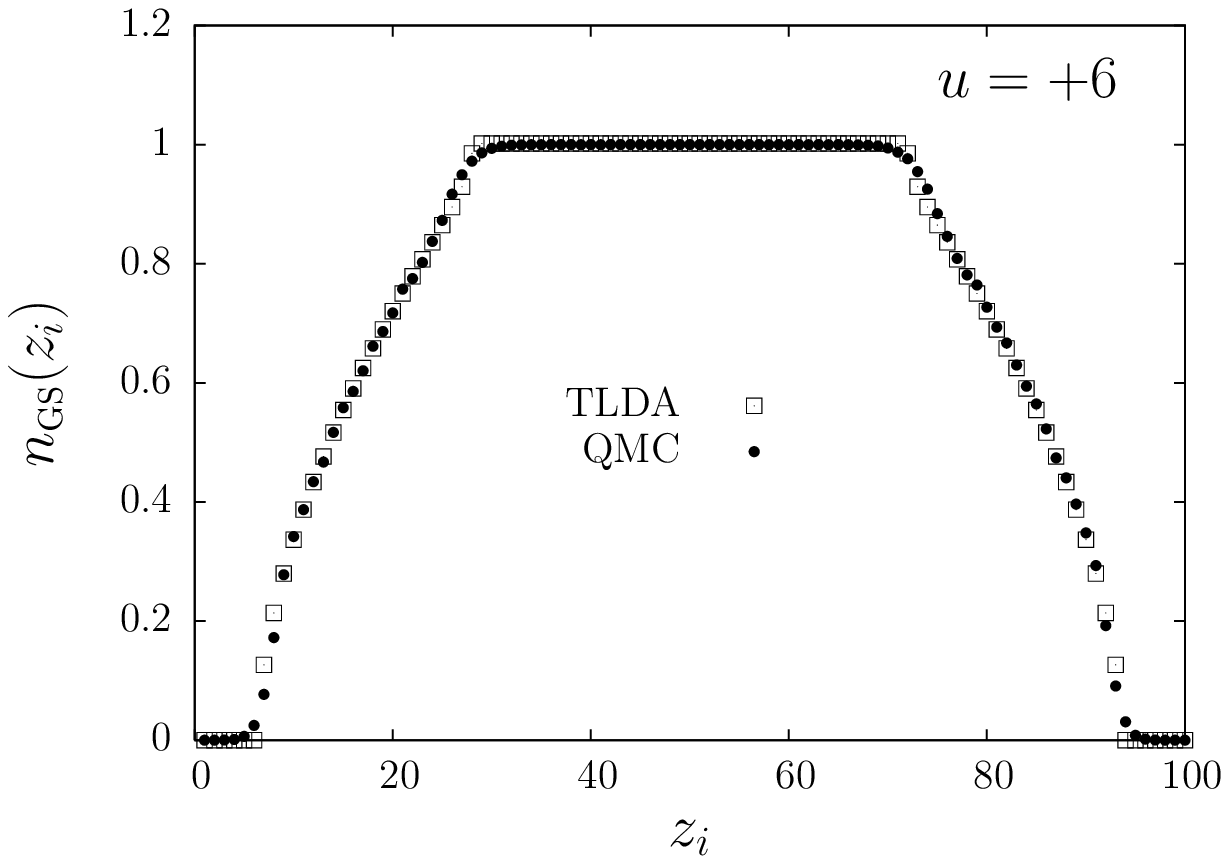}&
\includegraphics[width=0.50\linewidth]{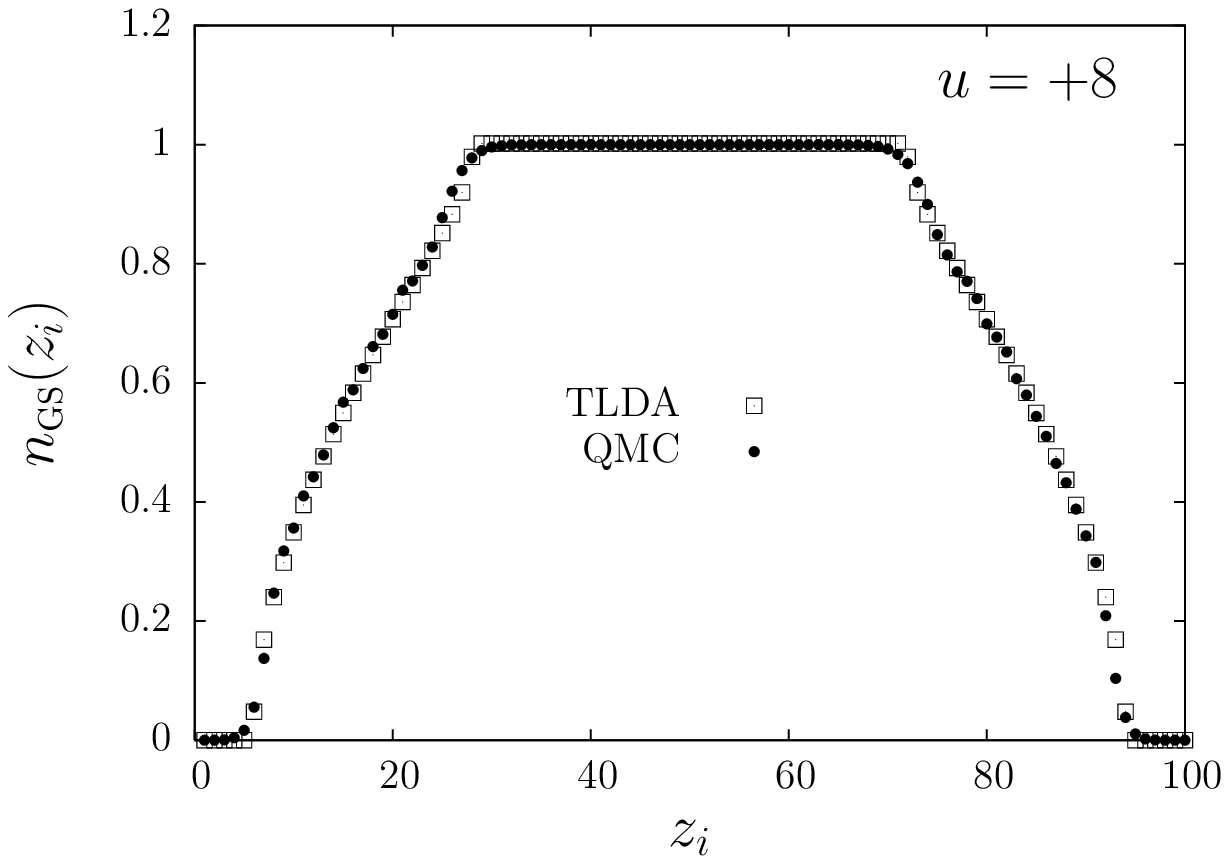}
\end{tabular}
\caption{Site occupation $n_{\rm \scriptscriptstyle GS}(z_i)$ as a function of $z_i$ 
for a system of $N_{\rm f}=70$ fermions with $u=+2,+4,+6$, and $+8$ in a lattice with $N_{\rm s}=100$ sites, confined by a harmonic potential of strength $V_2/t=2.5\times 10^{-3}$. 
${\rm BALDA}/{\rm FN}$ and TLDA results are compared with QMC data.\label{fig:eleven}}
\end{center}
\end{figure*}

We now turn to illustrate our main numerical results for the density profile of paramagnetic Fermi gases, which are summarized in Figs.~\ref{fig:three}-\ref{fig:eleven}. 
In this work we focus on external potentials of the general form~\cite{rigol_pra} 
\begin{equation}\label{eq:ext_pot}
V_{\rm ext}(z_i)=\sum_{\ell\geq 1}\, V_\ell\,(z_i-N_{\rm s}/2)^\ell\,,
\end{equation}
where $\ell$ is an integer and $V_\ell$ a constant.

We have solved numerically the self-consistent scheme represented by Eqs.~(\ref{eq:sks})-(\ref{eq:balda}) by using both the ${\rm BALDA}/{\rm LSOC}$ parametrization~\cite{lima_2003} and the ${\rm BALDA}/{\rm FN}$ procedure. 
Results obtained through the TLDA in Eqs.~(\ref{eq:tlda}) and~(\ref{eq:tf_scheme}) have also been examined. In parallel, QMC simulations have been performed using a zero-temperature 
projector approach~\cite{sugiyama_1986,sorella_1988} adapted from the QMC 
determinantal algorithm of Scalapino {\it et al.}~\cite{scalapino_1981}. 
Within this approach a projector operator $\exp(-\theta {\hat {\cal H}})$ 
is applied to a trial wave function, which we choose to be the exact 
ground state for $u=0$ (we have adopted a projector parameter $\theta$ with values up to $45/t$ for the accurate comparisons that we present in this work). 
We have used a discrete Hubbard-Stratonovich transformation~\cite{hirsch_1983} in decoupling the fermionic degrees of freedom in the interaction term of ${\hat {\cal H}}$. 
A detailed description of our QMC approach can be found in 
Refs.~\onlinecite{loh_1992,muramatsu_1999,assaad_2002}.

In Fig.~\ref{fig:three} we show the GS site occupation 
for a Fermi gas with $N_{\rm f}=30$ atoms trapped in a purely harmonic potential ({\it i.e.} $V_{\ell\neq 2}=0$) of strength $V_2/t=6\times 10^{-3}$ and in an optical lattice with $N_{\rm s}=100$ sites. 
The interaction parameter is increased from $u=+2$ to $u=+8$. 
The agreement between the ${\rm BALDA}/{\rm FN}$ scheme and 
the QMC results is clearly excellent for all values of $u$. 
The ${\rm BALDA}/{\rm LSOC}$ scheme
gives similar results and improves with increasing coupling ($u\gtrsim 6$), 
but overall the ${\rm BALDA}/{\rm FN}$ results are closer to the QMC data. In 
the rest of the paper we will thus focus on the ${\rm BALDA}/{\rm FN}$ scheme.

Computationally, ${\rm BALDA}/{\rm FN}$ is slightly more expensive than ${\rm BALDA}/{\rm LSOC}$. Both typically take a few seconds on a small PC to generate a single density profile like one of those reported in Fig.~\ref{fig:three}. 
QMC runs may take about $30$ hours for a single density profile, but unlike LDA also provide access to correlation functions and to the momentum distribution. In particular, the calculation of the GS energy, which is straightforward within the ${\rm BALDA}$-DFT scheme, requires a careful extrapolation procedure within QMC, which is explained in detail in Appendix~\ref{appendix_QMC} and illustrated in Fig.~\ref{fig:four}. The extrapolated QMC GS energies corresponding to the GS density profiles shown in Fig.~\ref{fig:three} have been reported in Table~\ref{table:one} with their estimated statistical error, together with the ${\rm BALDA}/{\rm FN}$ and ${\rm BALDA}/{\rm LSOC}$ results.

\begin{table}
\caption{Ground-state energy (in units of $t$) of a repulsive Fermi gas with $N_{\rm f}=30$ atoms, trapped in a harmonic potential of strength $V_2/t=6\times 10^{-3}$ and in a lattice with $N_{\rm s}=100$ sites.\label{table:one}}
\begin{ruledtabular}
\begin{tabular}{cccc}
$u$ & ${\rm BALDA/LSOC}$ & ${\rm BALDA/FN}$ & ${\rm QMC}$\\\hline\hline
$+2$  & $-17.16$  & $-16.68$ & $-16.68\pm0.07$ \\\hline
$+4$  & $-11.33$  & $-10.84$ & $-10.74\pm0.08$ \\\hline
$+6$  & $-8.00$   & $-7.52$  & $-7.64\pm0.08$  \\\hline
$+8$  & $-5.95$   & $-5.50$  & $-5.62\pm0.11$  \\ 
\end{tabular}
\end{ruledtabular}
\end{table}

In Fig.~\ref{fig:five} we compare TLDA results with ${\rm BALDA}/{\rm FN}$ and QMC results for the same system parameters as in Fig.~\ref{fig:three}, for the 
cases $u=+2$ and $u=+8$. TLDA is slightly less accurate than ${\rm BALDA}/{\rm FN}$, especially at relatively small values of the interaction parameter where hopping kinetic processes, which are treated at a simple LDA level in the TLDA, are still important. For the same reason, the regions close to the edge of the trap are those where the TLDA is less accurate.

In Fig.~\ref{fig:six} we show the local inverse compressibility,
\begin{equation}\label{eq:local_compressibility}
\kappa^{-1}(z_i)=\left.\frac{\partial^2 \varepsilon_{\rm \scriptscriptstyle GS}(n,u)}{\partial n^2}\right|_{n\rightarrow n_{\rm \scriptscriptstyle GS}(z_i)}\,,
\end{equation}
corresponding to the system parameters as in Fig.~\ref{fig:three}. This quantity is clearly much more sensitive to the variations of the interaction parameter $u$ than the GS site occupation.

In Fig.~\ref{fig:seven} we test the performance of the ${\rm BALDA}/{\rm FN}$ scheme
upon variations of the strength of the harmonic potential $V_2/t$ (top panel) and of the particle number $N_{\rm f}$ (bottom panel). Again, the agreement between the ${\rm BALDA}/{\rm FN}$ scheme and the QMC results is excellent for all the values that we have checked.

We show next some results for the GS density profiles of Fermi gases trapped in anharmonic potentials. 
Fig.~\ref{fig:eight} illustrates two possible situations. In the top panel we show an asymmetric external potential which contains a main harmonic component with strength $V_2/t=1.6\times 10^{-2}$, a cubic component with strength $V_3/t=1.6\times 10^{-4}$ breaking inversion symmetry, and a quartic component with strength $V_4/t=1.92\times 10^{-5}$ ensuring existence of a ground state. The presence of cubic and quartic components represents 
small deviations from a purely harmonic trapping potential that may occur in a real trap in the laboratory. In the bottom panel of Fig.~\ref{fig:eight} we show a double-well potential similar to the one that is used to model tunneling-coupled lateral semiconductor quantum dots: this potential contains a harmonic component with strength 
$V_2/t=-4\times 10^{-3}$ and a quartic component with strength $V_4/t=3\times10^{-6}$.

In the top panel of Fig.~\ref{fig:nine} we show the ${\rm BALDA}/{\rm FN}$ predictions and the QMC results for the 
GS site occupation of a Fermi gas with $N_{\rm f}=14$ atoms and $u=+8$, trapped in the asymmetric potential depicted in the top panel of Fig.~\ref{fig:eight}. In the same figure TLDA results 
are also shown. The performance of the TLDA scheme at weaker interactions 
deteriorates with decreasing particle number. 
In the bottom panel of Fig.~\ref{fig:nine} we show the GS site occupation for a Fermi gas with $u=+2$ trapped in the double-well potential depicted in the bottom panel of Fig.~\ref{fig:eight}.
In Fig.~\ref{fig:ten} we show the extension of Fig.~\ref{fig:nine} to the case of attractive interactions, which can be handled by means of the technique described in Ref.~\onlinecite{gao_2005}. The site occupation combines features found in Refs.~\onlinecite{gao_2005,continuous_systems} for attractive interactions in a single parabolic well, {\it i.e.} atomic-density waves induced by Luther-Emery spin pairing, with features shown in the bottom panel of Fig.~\ref{fig:nine} for a double well and repulsive interactions. Repulsive interactions are seen to lead to a higher density at the center of the double-well potential, suggesting enhanced tunneling between the wells.

In all these cases the ${\rm BALDA}/{\rm FN}$ scheme has been found to be extremely accurate, as judged by comparison to QMC. However, all data shown in Figs.~\ref{fig:three}-\ref{fig:ten} correspond to ``purely metallic" phases of the interacting Fermi gas, {\it i.e.} phases in which $n_{\rm \scriptscriptstyle GS}(z_i)<1$ everywhere inside the trap. From earlier work~\cite{rigol_prl,rigol_pra,xia_ji_2005} we know that in the trap there can be metallic phases that coexist with Mott-insulating and/or band-insulating regions, {\it i.e.} phases in which $n_{\rm \scriptscriptstyle GS}(z_i)$ is locally locked to $1$ or $2$. 
What happens with the ${\rm BALDA}/{\rm FN}$ scheme if the system parameters are such that $n_{\rm \scriptscriptstyle GS}(z_i)$ becomes unity or very close to unity at some point in space? As we have mentioned in Sect.~\ref{soft}, it is difficult in this case to obtain converged self-consistent solutions of the KS equations due to the discontinuity at $n=1$ in the xc potential of the $1D$ HHM. It is still possible to obtain a converged self-consistent solution in a very reasonable number of iterations if the discontinuity is relatively small. 
An example is given in the top left panel of Fig.~\ref{fig:eleven}, where 
we show the GS site occupation for a Fermi gas with $N_{\rm f}=70$ atoms
in an optical lattice with $N_{\rm s}=100$ sites, subject to a purely 
harmonic potential of strength $V_2/t=2.5\times 10^{-3}$. For $u=+2$ the size of the xc discontinuity is so 
small that we are able to achieve an accurate self-consistent solution of the ${\rm BALDA}/{\rm FN}$ equation. 
Upon further increasing $u$ in the same system while keeping unchanged all other control parameters, it becomes progressively difficult to obtain converged self-consistent solutions of the KS 
equations in a reasonable number of iterations. 
The easiest way to sidestep this problem is to resort to the TLDA scheme, Eqs.~(\ref{eq:tlda}) and~(\ref{eq:tf_scheme}), which is able to capture the main physical features of the above mentioned coexistence of compressible and incompressible regions~\cite{vivaldo_klaus_2005,ferconi_1995}. 

In Fig.~\ref{fig:eleven} we compare TLDA results with QMC results in the 
metal-insulator phase-separated regime. At $u=+2$ BALDA shows first indications
of a locally incompressible region (a plateau in the density profile at $n=1$),
whereas QMC data still predict the system to be fully metallic (compressible).
At larger $u$, the plateau also develops in the QMC calculations, but, as
explained in Sec.~\ref{baldasubsubsec} BALDA now ceases to converge. TLDA
calculations, on the other hand, are still possible. For fully developed
phase separation (cases of $u=+6$ and $u=+8$ in Fig.~\ref{fig:eleven}) 
TLDA and QMC agree very well.
At $u= +3$, the plateau at $n=1$ is more pronounced in TLDA than in QMC, where
the density profile just begins to flatten. In this particular case we have 
also performed a TLDA calculation using the analytical approach of Ref.~\onlinecite{vivaldo_klaus_2005}. 
Corresponding data are labelled by ${\rm TLDA}/{\rm LSOC}$ in
Fig.~\ref{fig:eleven}. For low densities (at the edges of the trap) fully
numerical TLDA data agree better with QMC, but in the center of the trap
and at the plateaus ${\rm TLDA}/{\rm LSOC}$ and QMC data agree better. For the GS energy at this particular value of $u$, ${\rm TLDA}$ gives ${\cal E}^{\rm \scriptscriptstyle TLDA}_{\rm \scriptscriptstyle GS}=16.57 t$ while ${\rm TLDA}/{\rm LSOC}$ gives ${\cal E}^{\rm \scriptscriptstyle TLDA/LSOC}_{\rm \scriptscriptstyle GS}=15.95 t$. These numbers have to be compared with the QMC value ${\cal E}^{\rm \scriptscriptstyle QMC}_{\rm \scriptscriptstyle GS}=(16.64\pm 0.09)t$.

Using the TF-like TLDA equation~(\ref{eq:tf_scheme}) a simple explanation for the formation of incompressible regions inside the trap in coexistence with metallic regions can be given. When the local density reaches unity, {\it i.e.} the value associated with the xc Mott gap, the left-hand side of Eq.~(\ref{eq:tf_scheme}) takes up the discontinuity $\Delta_{\rm xc}(u)$. This implies that the density resists crossing unity and develops instead an incompressible region where the constant value $n_{\rm \scriptscriptstyle GS}(z_i)=1$ is maintained up to a width $W$ such that the difference in the classical potential $U n_{\rm \scriptscriptstyle GS}(z_i)/2+V_{\rm ext}(z_i)$, evaluated at the end points of the incompressible region, exactly compensates for $\Delta_{\rm xc}(u)$. This criterion allows one to find {\it a priori} those regions of the trap where the local Mott-insulating incompressible phases are energetically favorable over metallic phases~\cite{vivaldo_klaus_2005}.

\section{Conclusions}
\label{discussion_conclusions}

In this work we have shown how a detailed picture of ground-state properties of strongly interacting $1D$ ultracold Fermi gases emerges through a novel DFT scheme using as reference fluid a many-body interacting $1D$ model which is exactly solvable, the $1D$ homogeneous Hubbard model. Needless to say, this basic idea is applicable to all inhomogeneous $1D$ systems for which a properly chosen, underlying homogeneous reference liquid can be described by an exactly solvable model, only the ground-state energy of the liquid being the key input~\cite{soft,lima_2003,burke_2004,hemo,bakhtiari_2005,continuous_systems}.

Our main conclusions are summarized as follows:

(i)	
Bethe-Ansatz-based density-functional calculations agree quantitatively with independent quantum Monte Carlo data. Typical differences between ${\rm BALDA}/{\rm FN}$ and QMC results for the density profiles in the metallic regime are around $1\%$. In the worst case encountered in all the calculations performed in this work the difference amounts to $7\%$. These errors are not easily reduced, but are sufficiently small to allow a detailed microscopic description of the energetics and the atom-density profiles of confined fermions, at a computational cost reduced by orders-of-magnitude compared to QMC. Large systems of thousands of sites and complex systems, {\it e.g.} with reduced spatial symmetries, are easily handled by the BALDA-DFT scheme. 

(ii) 
A fully numerical formulation of the BALDA requires solution of the Lieb-Wu integral equations instead of a parametrization of the energy of the reference fluid. The added numerical effort is compensated by a relatively small gain in accuracy, as judged by comparison of ${\rm BALDA}/{\rm LSOC}$ and ${\rm BALDA}/{\rm FN}$ density profiles to the QMC profiles. 

(iii)	
A total-energy LDA can be set up in the same way and provides useful numerical 
results in situations where Kohn-Sham calculations using BALDA for the
correlation energy fail to converge. These situations arise in the presence 
of a local Mott metal-insulator transition, leading
to phase separation characterized by flat (incompressible) and metallic
(compressible) regions in the density profiles.  

(iv) Overall, all four density-functional schemes (Kohn-Sham BALDA/FN and 
BALDA/LSOC where they converge, TLDA, and TLDA/LSOC) reveal the same 
physics also seen in the QMC data. DFT schemes, however, tend to predict
phase separation (plateau formation) at slightly lower values of the 
interaction $U$.

(v) 
Density profiles corresponding to different interaction strengths, filling factors in the metallic regime, and curvatures of the confining potential are quite similar. The local compressibility, on the other hand, depends sensitively on the details of the system, and can be used to discriminate between different choices of system parameters (see Fig.~\ref{fig:six}).

(vi)	
Different external potentials, such as asymmetric trappings and double-well structures, are easily handled by the same techniques. Specifically for the double well, a signature of tunneling between the two wells is a tilted density profile, which shows that atoms can tunnel from one well to the other even for very low filling factors. For attractive interactions, additionally density waves form separately in each well~\cite{gao_2005,continuous_systems}, until the density becomes so high that oscillations arising from both wells start to merge and a joint pattern develops. This type of double-well structure has, to our knowledge, not yet been produced optically in $1D$ ultracold atom systems, but is readily created in higher-dimensional traps and semiconductor heterostructures, to which many of our conclusions also apply.

\begin{acknowledgments}
M.P. acknowledges G. Vignale for many illuminating discussions on 
density-functional calculations of the edge structure of fractional 
quantum Hall systems. 
We also acknowledge useful discussions with P. Capuzzi, H. Hu, A. Kocharian, 
X.-J. Liu, E. Papa, V. Pellegrini, and S. Roddaro.
K.C. was supported by FAPESP and CNPq.
M.R. was supported by NFS-DMR-0312261, NFS-DMR-0240918, SFB 382, and 
HLR-Stuttgart (where most of the QMC calculations were done).
M.R. acknowledges A. Muramatsu and S. Wessel for insightful discussions.
\end{acknowledgments}

\appendix

\section{Some details of SOFT}
\label{appendix_SOFT}

The aim of this Appendix is to present a summary of the two key results of 
SOFT \cite{soft}: (i) the Hohenberg-Kohn theorem and (ii) the Kohn-Sham 
mapping to an auxiliary noninteracting system. 

The basic variable of SOFT is the site-occupation $n(z_i)=\langle\Psi|{\hat n}(z_i)|\Psi\rangle$, 
where $|\Psi\rangle$ is a generic many-body state. As in standard DFT, the central result of SOFT is the Hohenberg-Kohn (HK) theorem, which can be summarized in three key statements: (a) the GS expectation value of any observable ${\hat {\cal O}}$ is a unique functional ${\cal O}=\langle{\rm GS}|{\hat {\cal O}}|{\rm GS}\rangle={\cal O}[n_{\rm \scriptscriptstyle GS}]$ of the GS site-occupation $n_{\rm \scriptscriptstyle GS}(z_i)$; 
(b) the GS site-occupation minimizes the total-energy functional ${\cal E}[n]$; and (c) 
${\cal E}[n]$ can be written as
\begin{equation}\label{eq:totenergy}
{\cal E}[n]={\cal F}_{\rm HK}[n]+\sum_{i}V_{\rm ext}(z_i)n(z_i)\,,
\end{equation}
where ${\cal F}_{\rm HK}[n]=\langle\Psi|{\hat {\cal T}}+{\hat {\cal H}}_{\rm int}|\Psi\rangle$ is a 
{\it universal} functional of the site occupation, in the sense that it does not depend on the external potential. 

Part (b) of the HK theorem suggests that if the exact analytical expression of ${\cal F}_{\rm HK}[n]$ was known, the GS energy and the GS site occupation could be found by solving the Euler-Lagrange equation
\begin{equation}\label{eq:HK_variational}
\frac{\delta {\cal F}_{\rm HK}[n]}{\delta n(z_i)}+V_{\rm ext}(z_i)={\rm constant}\,,
\end{equation}
the constant having the meaning of a Lagrange multiplier to enforce particle-number conservation.

The Kohn-Sham mapping, again in analogy with standard DFT, provides an essential simplification. One considers a noninteracting auxiliary system described by the Hamiltonian
\begin{eqnarray}
{\hat {\cal H}}_{\rm s}&=&-\sum_{i,j}\sum_\sigma t_{i,j}
\left[{\hat c}^{\dagger}_{\sigma}(z_i){\hat c}_{\sigma}(z_j)+{\rm H}.{\rm c}.\right]\nonumber\\
&+&\sum_{i}v_{\rm \scriptscriptstyle KS}(z_i){\hat n}(z_i)\,.
\end{eqnarray}
The central assertion used in establishing the mapping is that for any interacting system there exists a local single-particle potential $v_{\rm \scriptscriptstyle KS}(z_i)$ such that the exact GS site occupation $n_{\rm \scriptscriptstyle GS}(z_i)$ of the interacting system equals the GS site occupation of the auxiliary problem $n_{\rm \scriptscriptstyle GS}(z_i)=n^{(s)}_{\rm \scriptscriptstyle GS}(z_i)$ (noninteracting $v$-representability). According to part (c) of the HK theorem there then exists a unique energy functional
${\cal E}_{\rm s}[n]={\cal T}_{\rm s}[n]+\sum_{i}v_{\rm \scriptscriptstyle KS}(z_i)n(z_i)$, 
for which the variational equation $\delta {\cal E}_{\rm s}[n]=0$ yields the exact GS site occupation $n^{(s)}_{\rm \scriptscriptstyle GS}(z_i)$ corresponding to ${\hat {\cal H}}_{\rm s}$. ${\cal T}_{\rm s}[n]$ denotes the universal kinetic energy functional of noninteracting pseudospin-$1/2$ fermions.

Suppose that the ground state of ${\hat {\cal H}}_{\rm s}$ is nondegenerate. 
The GS site occupation $n^{(s)}_{\rm \scriptscriptstyle GS}(z_i)$ 
(and thus, by assumption, $n_{\rm \scriptscriptstyle GS}(z_i)$) possesses a unique representation
\begin{equation}\label{eq:gsdensity}
n_{\rm \scriptscriptstyle GS}(z_i)=\sum_{\alpha, {\rm occ.}} |\varphi_\alpha(z_i)|^2
\end{equation}
in terms of the lowest $N_{\rm f}$ single-particle orbitals obtained from the 
KS-Schr\"odinger equation
\begin{equation}\label{eq:ksequations}
\sum_j \left[-t_{i,j}+v_{\rm \scriptscriptstyle KS}(z_i)\delta_{ij}\right]\varphi_\alpha(z_j)=\varepsilon_\alpha \varphi_\alpha(z_i)\,.
\end{equation}
Once the existence of a potential $v_{\rm \scriptscriptstyle KS}(z_i)$ generating $n_{\rm \scriptscriptstyle GS}(z_i)$ {\it via} Eqs.~(\ref{eq:gsdensity}) and (\ref{eq:ksequations}) is assumed, uniqueness of $v_{\rm \scriptscriptstyle KS}(z_i)$ follows from the HK theorem. Thus the single-particle orbitals $\varphi_\alpha(z_i)=\varphi_\alpha[n_{\rm \scriptscriptstyle GS}](z_i)$ are unique functionals of $n_{\rm \scriptscriptstyle GS}(z_i)$, and the noninteracting kinetic energy 
\begin{equation}\label{eq:noninteracting_ke_functional}
{\cal T}_{\rm s}[n_{\rm \scriptscriptstyle GS}]=-\sum_{i,j}\sum_\alpha t_{i,j}\left[\varphi^\star_\alpha(z_i)\varphi_\alpha(z_j)+{\rm c}.{\rm c}.\right]
\end{equation}
is a unique functional of $n_{\rm \scriptscriptstyle GS}(z_i)$ as well.

It is convenient at this point to write the total energy functional ${\cal E}_{\rm \scriptscriptstyle GS}[n_{\rm \scriptscriptstyle GS}]$ in Eq.~(\ref{eq:totenergy}) by adding and subtracting ${\cal T}_{\rm s}[n_{\rm \scriptscriptstyle GS}]$ and a Hartree term ${\cal E}_{\rm H}=U\sum_i n^2_{\rm \scriptscriptstyle GS}(z_i)/4$, 
{\it i.e.}
\begin{eqnarray}
{\cal E}_{\rm \scriptscriptstyle GS}[n_{\rm \scriptscriptstyle GS}]&=&{\cal T}_{\rm s}[n_{\rm \scriptscriptstyle GS}]+{\cal E}_{\rm H}[n_{\rm \scriptscriptstyle GS}]+{\cal E}_{\rm xc}[n_{\rm \scriptscriptstyle GS}]\nonumber\\
&+&\sum_{i}V_{\rm ext}(z_i)n_{\rm \scriptscriptstyle GS}(z_i)\,,
\end{eqnarray}
where the exchange-correlation functional is formally defined as ${\cal E}_{\rm xc}[n_{\rm \scriptscriptstyle GS}]\equiv{\cal F}_{\rm HK}[n_{\rm \scriptscriptstyle GS}]-{\cal T}_{\rm s}[n_{\rm \scriptscriptstyle GS}]-{\cal E}_{\rm H}[n_{\rm \scriptscriptstyle GS}]$. The HK variational principle ensures that ${\cal E}[n]$ is stationary for small variations $\delta n(z_i)$ around $n_{\rm \scriptscriptstyle GS}(z_i)$:
\begin{widetext}
\begin{equation}
{\cal E}[n_{\rm \scriptscriptstyle GS}+\delta n]-{\cal E}_{\rm \scriptscriptstyle GS}[n_{\rm \scriptscriptstyle GS}]=\delta {\cal T}_{\rm s}+\sum_i \delta n(z_i)\left[V_{\rm ext}(z_i)+\frac{U}{2}n_{\rm \scriptscriptstyle GS}(z_i)+v_{\rm xc}[n_{\rm \scriptscriptstyle GS}](z_i)\right]=0\,,
\end{equation}
\end{widetext}
where $v_{\rm xc}[n_{\rm \scriptscriptstyle GS}](z_i)$ denotes the exchange-correlation potential,
\begin{equation}\label{eq:formalvxc}
v_{\rm xc}[n_{\rm \scriptscriptstyle GS}](z_i)=\left.\frac{\delta {\cal E}_{\rm xc}[n]}{\delta n(z_i)}\right|_{n_{\rm \scriptscriptstyle GS}(z_i)}\,.
\end{equation}
Using $\delta {\cal T}_{\rm s}=-\sum_{i}v_{\rm \scriptscriptstyle KS}(z_i)\delta n(z_i)$, we find that the Kohn-Sham potential is given by 
\begin{equation}
v_{\rm \scriptscriptstyle KS}(z_i)=v_{\rm H}[n_{\rm \scriptscriptstyle GS}](z_i)+v_{\rm xc}[n_{\rm \scriptscriptstyle GS}](z_i)+V_{\rm ext}(z_i)\,,
\end{equation}
where $v_{\rm H}[n_{\rm \scriptscriptstyle GS}](z_i)=Un_{\rm \scriptscriptstyle GS}(z_i)/2$.

\section{Quantum Monte Carlo calculation of the ground-state energy}
\label{appendix_QMC}

Within our zero-temperature QMC approach, we use the Trotter 
decomposition~\cite{fye_1986} in applying the projector operator to a trial wave function $|\Psi_{\rm T}\rangle$, 
\begin{equation}
\exp(-\theta\,{\hat {\cal H}})\,|\Psi_{\rm T}\rangle=
\left[ \exp(-\Delta \tau\,{\hat {\cal H}})\right]^{\theta/\Delta \tau}\,|\Psi_{\rm T}\rangle\,.
\end{equation} 
For small values of $\Delta \tau$ one can 
then split the exponential of a sum of noncommuting operators as~\cite{fye_1986} 
\begin{eqnarray}
\exp(-\Delta\tau\,{\hat {\cal H}})&=&\exp(-\Delta\tau\,{\hat {\cal T}})
\exp(-\Delta\tau\,{\hat {\cal H}}_{\rm int})\nonumber\\
&\times&\exp(-\Delta\tau\,{\hat {\cal H}}_{\rm ext})+{\cal O}((\Delta \tau)^2)\,.
\end{eqnarray}
This is essential for the implementation of the determinantal QMC algorithm~\cite{scalapino_1981,loh_1992,muramatsu_1999,assaad_2002}.

The error introduced by the Trotter decomposition in the calculation of 
the ground-state energy can be shown to be of order 
$(\Delta \tau)^2$~\cite{assaad_2002}. Hence, for our comparison of the QMC 
energies with the ones of the ${\rm BALDA/FN}$ scheme 
we have made an extrapolation $\Delta \tau \rightarrow 0$. 
In Fig.~\ref{fig:four} we show the QMC ground-state energy as 
a function of $(\Delta \tau)^2$ and the linear fit ${\cal E}^{\rm \scriptscriptstyle QMC}_{\rm \scriptscriptstyle GS}(\Delta\tau)={\cal E}^{\rm \scriptscriptstyle QMC}_{\rm \scriptscriptstyle GS}+A (\Delta \tau)^2$ from which we get the QMC energies presented in Table~\ref{table:one}.

\end{document}